\documentclass[11pt]{article}
\usepackage{jheppub}

\usepackage[T1]{fontenc} 

\usepackage{amssymb}
\usepackage{amsmath}
\usepackage[title]{appendix}
\usepackage{graphicx}
\usepackage{graphics}
\usepackage{color}
\usepackage{fancyhdr}
\usepackage{graphicx}
\usepackage{float}
\usepackage{enumerate}
\usepackage{soul}

\usepackage[colorlinks=true,linktocpage=true,linkcolor=blue,citecolor=blue]{hyperref}

\usepackage[utf8]{inputenc}

    \newcommand{\be}{\begin{equation}}
  \newcommand{\ee}{\end{equation}}
    \newcommand{\ba}{\begin{align}}
  \newcommand{\ea}{\end{align}}

\definecolor{darkgreen}{RGB}{50,150,0}

\begin{document}

\title{Swampland Bounds on Dark Sectors}

\author[a]{Miguel Montero,}
\emailAdd{mmontero@g.harvard.edu}
\author[b]{Julian B.~Mu\~{n}oz,}
\emailAdd{julianmunoz@cfa.harvard.edu}
\author[a]{and Georges Obied}
\emailAdd{gobied@g.harvard.edu}

\affiliation[a]{ Department of Physics, Harvard University, Cambridge, MA 02138, USA}
\affiliation[b]{ Center for Astrophysics $\vert$ Harvard \& Smithsonian, 60 Garden St, Cambridge, MA, 02138, USA}

\abstract{
We use Swampland principles to theoretically disfavor regions of the parameter space of dark matter and other darkly charged particles that may exist.
The Festina Lente bound, the analogue of the Weak-Gravity conjecture in de Sitter, places constraints on the mass and charge of dark particles, which here we show cover regions in parameter space that are currently allowed by observations.
As a consequence, a broad set of new ultra-light particles are in the Swampland, independently of their cosmic abundance, showing the complementarity of Quantum Gravity limits with laboratory and astrophysical studies.
In parallel, a Swampland bound on the UV cutoff associated to the axion giving a St\"{u}ckelberg photon its longitudinal mode translates to a new constraint on the kinetic mixings and masses of dark photons.
This covers part of the parameter space targeted by upcoming dark-photon direct-detection experiments.
Moreover, it puts astrophysically interesting models in the Swampland, including freeze-in dark matter through an ultra-light dark photon, as well as radio models invoked to explain the 21-cm EDGES anomaly.
}

\maketitle

\begin{section}{Introduction}

The particle content of our universe appears to contain a dark sector, which so far has eluded all direct probes of its nature.
This sector is composed of at least dark matter and dark energy, and thus requires new physics to be added to the---otherwise highly successful---Standard Model.
Moreover, different arguments ranging from data anomalies (e.g.,~\cite{Muong-2:2021ojo,LHCb:2021trn,cdf2022high}) to theoretical ones, such as the hierarchy problem and neutrino masses, demand for new low-energy physics to exist in our universe.
As such, the properties and nature of new dark sectors beyond the Standard Model is one of the key open questions in particle physics and cosmology.

The experimental program to detect such new physics is highly diverse, given the broad spectrum of possibilities.
For instance, dark-matter candidates range from ultra-light bosons (with masses as low as $m\sim 10^{-22}$ eV~\cite{Hui:2016ltb}) to super-Planckian objects, such as primordial black holes~\cite{Carr:2021bzv}, including the more traditional candidates at the weak scale~\cite{Roszkowski:2017nbc}.
Particle-physics experiments have been able to robustly test new physics up to energy scales $\Lambda\sim$ TeV, if their couplings are large enough to be produced in colliders. Direct-detection experiments, such as XENON \cite{XENON:2018voc} and LUX~\cite{LUX:2015abn}, are placing significant constraints on the WIMP paradigm \cite{Roszkowski:2017nbc}, which has encouraged the community to focus on lighter DM candidates not directly related to electroweak physics.
In line with this, cosmological and astrophysical observations are sensitive to new degrees of freedom at lower masses and energies, and can in principle reach much smaller couplings~\cite{Raffelt:1996wa,Alvarez:2014vva,CMB-S4:2016ple}.

In parallel, there have been recent advances in our understanding of the low-energy implications of quantum gravity (QG).
Naively, it seems difficult to connect the ``low-energy'' (i.e., sub-Planckian) world to QG.
For instance, it would appear that any effective quantum field theory (EFT) could be coupled to dynamical gravity.
However, there are self-consistent EFTs that can never arise as the low-energy limit of a quantum theory of gravity.
These EFT's are said to lie in ``the Swampland''~\cite{Vafa:2005ui} (as opposed to ``the Landscape''; see e.g.~\cite{Brennan:2017rbf,Palti:2019pca,vanBeest:2021lhn} for reviews).
Theories that can be placed robustly in the Swampland are, therefore, {\it theoretically} disfavored, as they cannot be consistently UV-completed when including gravity.

In this note, we open the study of the implications of this Swampland program for the dark sectors of our universe, and show its complementarity to both cosmological and experimental probes of new physics.

We will take advantage of recent progress in the Swampland literature, including the Festina Lente (FL) bound proposed in \cite{Montero:2019ekk} (an extension of the Weak-Gravity conjecture \cite{ArkaniHamed:2006dz} to de Sitter space) as well as the photon-mass bounds from~\cite{Reece:2018zvv},
to place new constraints on the existence of new dark sectors.
We consider vector-portal models (i.e., sectors with dark photons that may kinetically mix with ours), and study the cases of
\begin{enumerate}[i)]
\item  millicharged particles,
\item a secluded dark sector (with negligible kinetic mixing), and
\item new massive (but light) dark photons.
\end{enumerate}
In the first two cases we employ the FL bound to constrain new very light, charged particles, showing that part of the parameter space that will be probed by new experiments is in the Swampland.

In the dark-photon case we will use the conjectures in \cite{Reece:2018zvv}, as well as the generic mixing from~\cite{Obied:2021zjc}, to constrain dark-photon masses as well as their interactions.
One of the Swampland insights into phenomenology is that St\"{u}ckelberg masses require the existence of a radial mode.
We show that this radial mode $\sigma$ can be produced in astrophysical environments, rendering the St\"{u}ckelberg case similar to the Higgs one. In addition, the angular mode of a St\"{u}ckelberg photon is an axion and a Swampland bound on the UV cutoff of an axion EFT can be applied to models of the dark photon as well.
This allows us to place a portion of the light dark-photon parameter space (roughly those with masses $m_{A'} \lesssim 20\epsilon$ eV, given a kinetic mixing $\epsilon$) in the Swampland.

We will additionally briefly study how other models are in tension with the Swampland bounds, and how these bounds can interface with physics during inflation.
As we will show, the Swampland program can reach a broad set of models targeted by both dark-matter experiments and astrophysical observations.

We caution the reader that so far there are no universal proofs of these Swampland constraints (see e.g.~\cite{Cheung:2018cwt,Hamada:2018dde,Arkani-Hamed:2021ajd,Grimm:2019ixq} for recent efforts), as we lack a framework to prove general statements in QG.
Nevertheless, these constraints are supported by several different lines of evidence, coming from general arguments based on black-hole physics, unitarity, or String Theory (which provides a concrete model of quantum gravity in which we can test Swampland constraints). The following is a table of the Swampland principles that we use in this paper, together with a one-liner explanation of what expected property of quantum gravity ``would break'' if new physics was found in violation of each of the bounds.

\begin{table}[!h]
\begin{tabular}{c|c|c|c}
\textbf{Constraint} & \textbf{Statement} & \textbf{What goes wrong if untrue?} & \textbf{Reference} \\\hline\hline
Weak Gravity  (WGC) & $m\lesssim g M_{\rm Pl}$ & \begin{tabular}{@{}c@{}} Charged black holes\\ cannot evaporate while\\ remaining sub-extremal\end{tabular} & \cite{ArkaniHamed:2006dz,Harlow:2022gzl}\\\hline
Festina Lente (FL) & $m^2\gtrsim \sqrt{6} g\, M_{\rm Pl}\, H$ &
\begin{tabular}{@{}c@{}}Horizon-sized charged\\ black holes in dS  evaporate \\to pathological space-times
\end{tabular}
& \cite{Montero:2019ekk,Montero:2021otb}\\\hline
Magnetic WGC& $\Lambda_{\text{UV}}\lesssim \sqrt{f M_{\rm Pl}}$& \begin{tabular}{@{}c@{}}EFT cutoff coming from\\ tension of WGC strings\end{tabular} &\cite{ArkaniHamed:2006dz,Dolan:2017vmn,Hebecker:2017wsu,Heidenreich:2017sim}\\\hline
Species scale& $\Lambda_{\text{UV}}\lesssim g^{1/3} M_{\rm Pl}$&\begin{tabular}{@{}c@{}} Scale at which loops of\\ WGC states makes local\\ EFT break down\end{tabular}&\cite{Arkani-Hamed:2005zuc,Distler:2005hi,Dimopoulos:2005ac,Dvali:2007hz}
\end{tabular}
\label{tab:Swampland}
\end{table}

The statements above will be further detailed in the main text below whenever relevant.

In this work we will take an agnostic approach, and explore the phenomenological consequences of the Swampland bounds we consider (as well as possible model-building approaches to evading the bounds, whenever possible).

The paper is organized as follows. In Section~\ref{sec:ChargedParticles}, we study the implications of the FL bound to the case of new charged particles,  briefly introducing the FL bound and its extension to multiple $U(1)$'s.
Then, in Section ~\ref{sec:MassiveDarkPhotons} we study the related massive dark photon case.
In Section~\ref{sec:NAmodels} we review the implications of the FL bound for models with non-Abelian fields, and in Appendix~\ref{app:FLandInflation} we make some general comments on the compatibility of the bound with inflation. Appendix \ref{app:CCrelaxion} contains some general comments regarding the application of these ideas to the scenario of cosmological relaxation.
We conclude in Section \ref{sec:conclusions}.
Throughout the text we will briefly review the relevant Swampland bounds for the astroparticle reader, as well as the phenomenology of the models for the formal reader,
and will work in natural units.

\end{section}

\section{New Charged Particles}
\label{sec:ChargedParticles}

We begin by studying the case of particles charged under a new $U(1)$, i.e.,  millicharged and darkly charged particles.

\subsection{Formalism}

As is well known, a new (dark) photon can kinetically mix with its Standard Model counterpart.
The kinetic-mixing operator, being dimension 4, is not suppressed by a high-energy scale and can therefore leave an imprint at low energies despite its UV origin~\cite{Holdom:1985ag}, which makes it phenomenologically appealing (for a treatment of irrelevant portal interactions, see for e.g.~\cite{Contino:2020tix}).
For this reason, darkly charged particles are an excellent dark-matter candidate (e.g.~\cite{Agrawal:2016quu}), and  a plethora of experimental efforts are directed towards detecting them.
Here we will appeal to Festina Lente arguments (see the Table in the Introduction) to understand what regions of parameter space are in the Swampland and which others are favored experimental targets.

In order to set our notation, we start by considering a theory with an Einstein-Maxwell Lagrangian coupled to multiple $U(1)$ fields \cite{Holdom:1985ag,Berlin:2020pey}
\begin{equation}
S\supset\int d^4x\sqrt{-g}\left[ \frac{1}{16\pi G}\left(-R+2\Lambda\right)+ \frac{1}{4}\sum_{i=1,2}F_{i\mu\nu}F_i^{\mu\nu}
-\frac{\epsilon}{2}F_{1\mu\nu} F_2^{\mu\nu}
\right].
\label{lagg1}
\end{equation}
Here, $R$ and $\Lambda$ are the Einstein-Hilbert term and cosmological constant respectively, $\epsilon$ is the kinetic mixing parameter, and the index $i$ runs over the two $U(1)$'s, which in this Section will be assumed to be massless (we will lift this restriction in Sec.~\ref{sec:MassiveDarkPhotons}).
In that case, we can diagonalize the kinetic term by defining two new gauge bosons,
\begin{equation}
    \vec A = \begin{pmatrix}A \\A'\end{pmatrix} = \mathcal M \begin{pmatrix}A_1 \\A_2\end{pmatrix},
\end{equation}
which will be the regular photon ($A$) and the dark photon ($A'$).
We choose the matrix $\mathcal M$ to provide a diagonal kinetic term.
In that case, we have
\begin{equation}
S\supset\int d^4x\sqrt{-g}\left[ \frac{1}{16\pi G}\left(-R+2\Lambda\right)+ \frac{1}{4}F_{\mu\nu}F^{\mu\nu}+ \frac{1}{4}F'_{\mu\nu}F'^{\mu\nu}
\right].
\label{lagg2}
\end{equation}

In the absence of charged particles, this diagonalization is largely irrelevant.
However, these Lagrangians alone are not consistent with Swampland principles.
In particular, the Completeness Principle \cite{Polchinski:2003bq,Harlow:2018tng} requires that there are physical states with all possible charges. The Weak Gravity Conjecture, and its extension to multiple $U(1)$ fields, dubbed the ``Convex Hull Condition'' \cite{Cheung:2014vva}, can be regarded as stronger versions of the Completeness Principle that put additional constraints on the kinematic properties of particles (masses and charges). This relationship can be made precise \cite{Cordova:2022rer}.
We can satisfy this condition by including charged particles, in particular the electron of the Standard Model and an additional particle, a dark electron $\chi$, minimally coupled to the $A_2$ gauge field.

For the range of parameters of interest to this work, the regime $\epsilon\sim \mathcal{O}(1)$ is experimentally ruled out (see discussion below and in Figure~\ref{fig:MCP}). We will therefore work in the $\epsilon \ll 1$ approximation from this point onward.
In our normalization, a charged particle $i$ on a worldline $\mathcal{W}$ couples to the gauge fields $\vec{A}=(A,A')$ in~\eqref{lagg2} (i.e., the ones with diagonal kinetic terms) via
\begin{equation} \int_{\mathcal{W}} \mathcal {\vec {Q}}_i \cdot \vec{A}.\end{equation}
Here, the vector $\mathcal {\vec {Q}}_i$ belongs to some lattice $\Lambda_{\mathcal Q}$. We do not know what the full charge lattice is, but it must include at least a vector $\vec {\mathcal {Q}}_e$ for the electron. By choosing the appropriate basis (via an orthogonal transformation that preserves \eqref{lagg2}), we have ensured that it takes the form $\vec {\mathcal {Q}}_e = (e,0)$.
We are also assuming the model contains a millicharged particle $\chi$, with a charge vector parametrized as
\begin{equation}
\vec {\mathcal {Q}}_\chi = g'  \left(\epsilon, 1\right)
\label{delectron}
\end{equation}
in the same basis, where now $g'$ is the gauge coupling of the $\chi$ particle to the dark photon $A'$ in~\eqref{lagg2}.
As a consequence, $\chi$ aquires a millicharge under the SM photon~\cite{Holdom:1985ag}, of value $q_\chi = g'\epsilon/e$ (in units of the electron charge, as customary).

\subsection{The Festina Lente Bound}
\label{sec:FLbound}
With our notation set, we move on to apply the Festina Lente bound from Ref.~\cite{Montero:2019ekk}.
Reference \cite{Montero:2019ekk} studied evaporation of charged black holes in a de Sitter background, and by applying the rationale behind the successful Weak Gravity Conjecture (WGC) \cite{ArkaniHamed:2006dz}, proposed a constraint on the spectrum of charged particles, which applies to any $U(1)$ field in de Sitter space (with expansion rate $H$). This bound, the Festina Lente (FL) bound,  demands that for a minimally coupled $U(1)$, the spectrum of charged states satisfies
\begin{equation} m^2\geq \, \sqrt{6}\, g\, M_{\rm Pl} H\label{FL}\end{equation}
where $g$ is the $U(1)$ charge and $m$ its mass. Crucially, \eqref{FL} must be satisfied by all charged states in the theory. This bound is satisfied in the Standard Model today, where the quantity in the right-hand side of \eqref{FL} is roughly $M_{\rm Pl} H_0\sim \mathcal O(\rm meV^2)$, or around neutrino mass scale, whereas the lightest charged particle is the electron.

\begin{figure}[hbtp!]
\centering
	\includegraphics[width=0.4\textwidth]{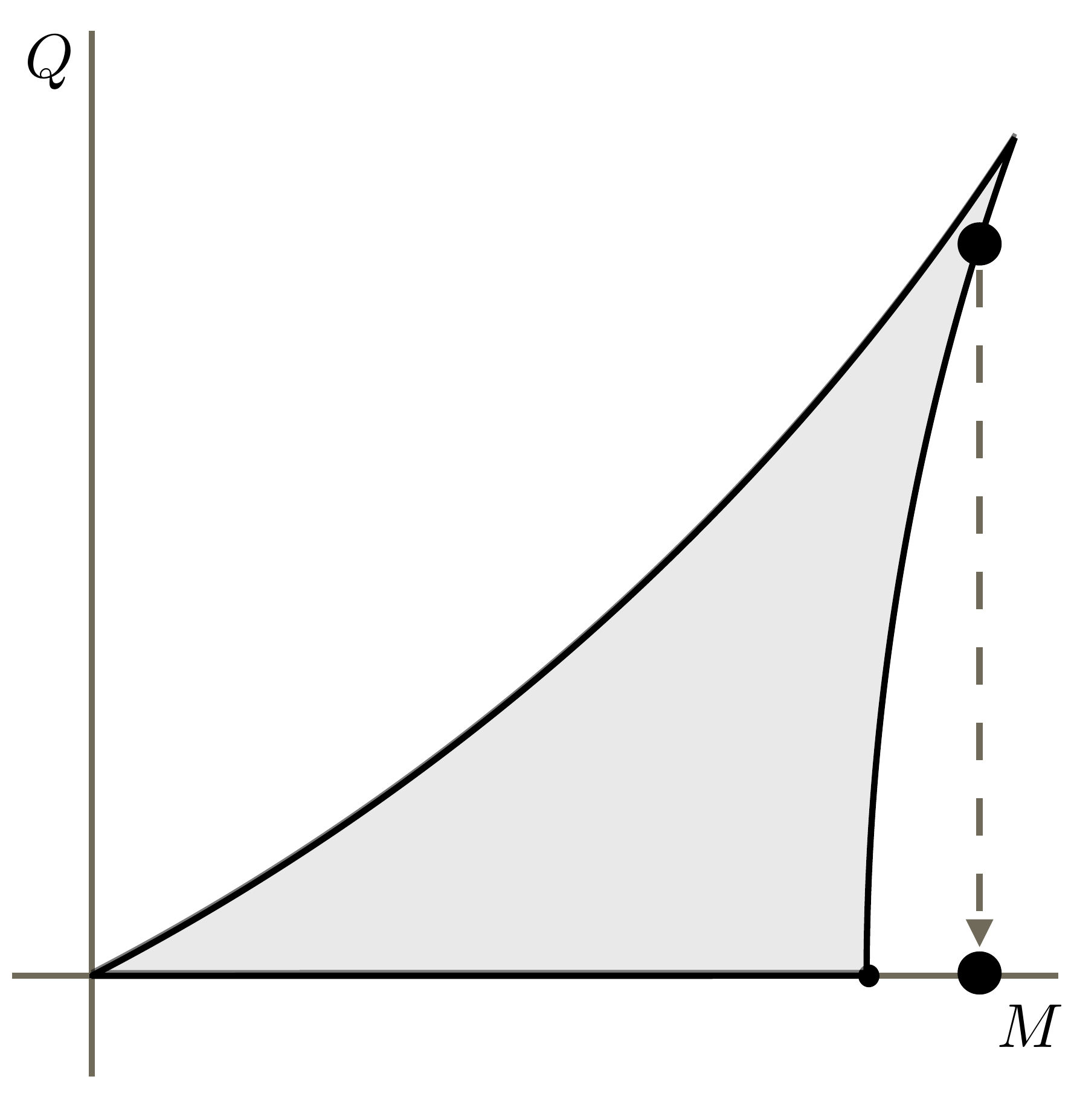}
	\caption{Charge versus mass plot for Reissner-Nordstrom-de Sitter black holes. Sub-extremal solutions (those without naked singularities) only exist inside the gray-shaded ``shark-fin''-shaped region. Unlike in flat space, there is a maximal value of the mass for a given value of the charge (the right-side edge of the shaded region); this corresponds to the so-called Nariai black hole, for which the cosmological and black-hole horizons coincide. A Nariai black hole with some charge $Q$ amd $M$ can decay following the dashed line if there exist particles in the spectrum that violate the FL bound,
	becoming super-extremal and thus
	producing a pathological spacetime that does not evaporate back to empty de Sitter space. The FL bound is the condition that these pathological decays do not ocurr. }
	\label{fig:FLDiagram}
\end{figure}

\begin{figure}[hbtp!]
\centering
	\includegraphics[width=0.8\textwidth]{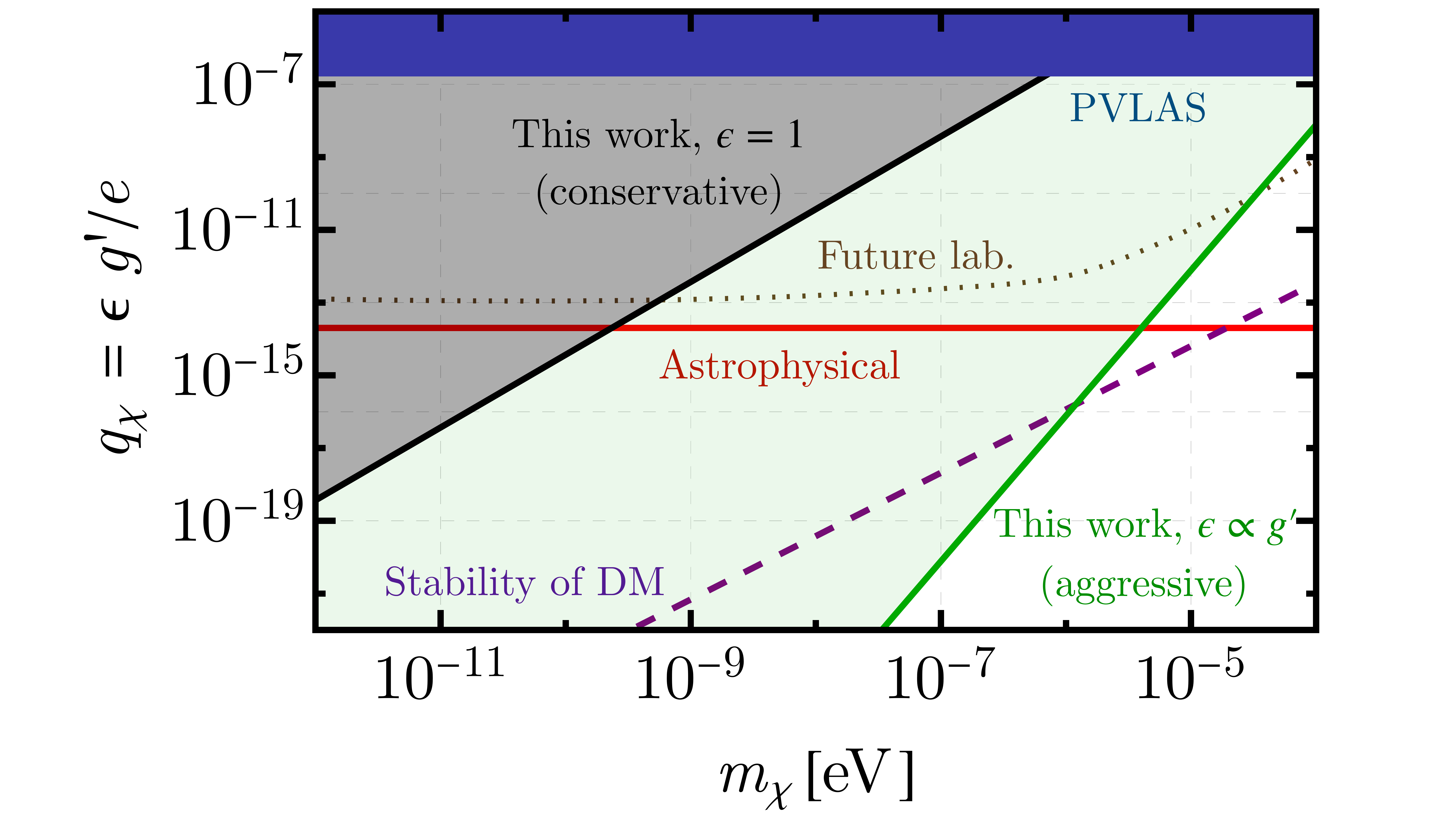}
	\caption{
Parameter space of millicharged particles (MCPs) of mass $m_\chi$ and charge $q_\chi$ in units of the electron charge.
	New particles in the blue region are ruled out by the PVLAS experiment~\cite{Ahlers:2006iz,DellaValle:2014xoa}.
	MCPs in the black-shaded region, from Eq.~\eqref{eq:qDMcons}, would violate the Festina-Lente bound, as they make black holes evaporate to pathological spacetimes in dS (see Fig.~\ref{fig:FLDiagram}).
	They are thus ``in the Swampland'', and theoretically disallowed.
	The region above the green line, given by Eq.~\eqref{eq:qDMagg} is  disfavored when further assuming that the kinetic mixing is $\epsilon = e\,g'/(16\pi^2)$, rather than just $\epsilon \leq 1$.
	The area above the red line is constrained by astrophysical observations of the tip of the red giant branch~\cite{Vogel:2013raa} (though this can be circumvented~\cite{Berlin:2020pey}), whereas above the purple dashed line there are limits related to its stability as DM~\cite{Jaeckel:2021xyo}.
	The dotted brown line is the forecasted sensitivity of a future laboratory experiment proposed in Ref.~\cite{Berlin:2020pey}.
	}
	\label{fig:MCP}
\end{figure}

The bound \eqref{FL} can be derived studying the decay of Reissner-Nordstrom-de Sitter black holes. Unlike in flat space, where one can have black holes of arbitrary mass and charge provided that $Q\leq M$ in Planck units, in de Sitter space there is a limit to both the mass and charge that a black hole can have. This maximally charged black hole, the Nariai black hole \cite{nariai1950some,Romans:1991nq}, is the largest black hole that fits within the cosmological horizon, as illustrated in Figure \ref{fig:FLDiagram}.

Much like their flat-space counterparts, black holes in de Sitter evaporate by slowly emitting charged particles, which slowly discharges the black hole . This is a quantum-mechanical process, which corresponds to Schwinger pair production of electrically charged particles in the near-horizon region of the black hole, where the electric field is strongest \cite{Hiscock:1990ex,Gibbons:1975kk,Schwinger:1951nm}. The Schwinger current has, schematically, a suppression factor
\begin{equation} \mathcal{J}\sim e^{-\frac{m^2}{qE}},\end{equation}
where $m$ is the mass of the charged particle being emitted, and $E$ is the near-horizon value of the black-hole electric field. If $m^2\ll qE$, the black hole will evaporate quickly; otherwise, the decay process is very slow.

 If the Schwinger current is not suppressed, the black hole will evaporate to a singular space time, instead of empty de Sitter space. Intuitively, if the current is unsuppressed, $m^2\ll qE$, then a black hole will lose charge much faster than it loses mass (see vertical line in Fig.~\ref{fig:FLDiagram}), effectively becoming overextremal, and leading to a singular spacetime; the black hole does not evaporate to empty de Sitter space. This is in tension with the principle, suggested by Weak Gravity, that every charged black hole should be able to evaporate back to empty space. This has been verified in every string compactification known to date (see e.g.~the review~\cite{Harlow:2022gzl}), and there is some evidence in holography \cite{Harlow:2018tng} and from analyticity and causality in flat space \cite{Hamada:2018dde,Arkani-Hamed:2005zuc}. The electric field of the Nariai black hole is $E\sim \sqrt{6} gM_{\text{Pl}} H$; imposing that the Schwinger current is suppressed, $m^2\ll qE$, leads to \eqref{FL}.

In order to study the consequences of FL for new darkly charged particles we must first generalize \eqref{FL} to a setup with multiple $U(1)$ gauge fields.
This was done in \cite{Montero:2021otb}, but since it plays a central role in our current work, we review its derivation here briefly. The FL bound \eqref{FL} can also be written as $m^2>g E$, where $E$ is the electric field of the Nariai black hole. At the Nariai limit, the electrostatic energy in the black hole is comparable with the vacuum energy itself, $M_{\rm Pl}^2H^2$. When there is more than one $U(1)$ field, this energy density can be distributed between the different components of the electric field.
The corresponding generalization is then simply,
\begin{equation} m^2\geq \sqrt{6}(\vec{\mathcal Q}\cdot\vec{u})\, M_{\rm Pl} H,
\label{FL20}
\end{equation}
for a unit vector $\vec{u}$.
Because this relation has to hold for any unit vector $\vec{u}$, by taking $\vec{u}\propto\vec{\mathcal Q}$ we get
\begin{equation} m^2\geq \sqrt{6} | \vec{\mathcal Q} |\, M_{\rm Pl} H
\end{equation}
This expression is the FL counterpart of the ``convex hull condition'' of \cite{Cheung:2014vva} for the WGC. We emphasize that, unlike multi-field generalizations of the the WGC, which can be satisfied by  a finite number of particles satisfying the WGC, the FL bound must be satisfied by \emph{every} state with the appropriate charges in the theory, for otherwise we could find a black hole that discharges too rapidly, becoming superextremal.
Applying the above expression to the dark electron (with mass $m_\chi$ and a charge vector given as in Eq.~\eqref{delectron}) we obtain
\begin{equation}
m_\chi^2\geq \sqrt{6} g'\, M_{\rm Pl} H, \quad {\rm or\ equivalently} \quad g'\leq \dfrac{m_\chi^2}{\sqrt{6}M_{\rm Pl} H},
\label{eq:limitgprime}
\end{equation}
to leading order in $\epsilon$.
This has direct consequences for very light charged particles, which we now explore.

\subsection{Case I: Millicharges}

First we focus on the case of millicharges.
The kinetic mixing between the two photons induces a millicharge on the new $\chi$ particles of size
$q_\chi = g' \epsilon/e < g'/e$.
As a consequence, the upper limit in Eq.~\eqref{eq:limitgprime} can be conservatively applied to $q_\chi e$ as well, resulting in the limits that we show in Fig.~\ref{fig:MCP} (labeled as conservative, as we have only required $\epsilon \leq 1$.)
This constraint places Milli-Charged Particles (MCPs) with
\begin{equation}
q_\chi= g'\epsilon/e \geq (m_\chi/1.6\; \rm meV)^2
\label{eq:qDMcons}
\end{equation}
in the Swampland, which as is clear from its mass and charge dependence will be most relevant for very light and weakly charged particles $\chi$.

In deriving the previous bound, we were agnostic about the size of the kinetic mixing parameter, as long as $\epsilon < 1$.
A commonly considered value is
\begin{equation}
\epsilon \approx \dfrac{e g'}{16 \pi^2},
\label{eq:oneLoopMixing}
\end{equation}
which corresponds (times an $\mathcal{O}(1)$ logarithmic factor) to the mixing induced by integrating out a very heavy particle, with unit charge under both $A$ and $A'$. A priori, taking \eqref{eq:oneLoopMixing} as an estimate for the kinetic mixing may seem unwarranted, since we do not know anything about the spectrum of massive states of the theory; in fact, the common situation in string theory is that one has infinite towers of states \cite{Ooguri:2006in}, with increasing values of the charges. On top of this, there could (theoretically) just be a bare kinetic mixing of order 1.

In spite of these caveats, it turns out that \eqref{eq:oneLoopMixing} is a good estimate for the magnitude of the kinetic mixing in a large class of perturbative string-theory models, as described in \cite{Obied:2021zjc}. This is because one-loop contributions of higher states in the tower cancel out. It is also in line with the emergence proposal of \cite{Heidenreich:2017sim,Grimm:2018ohb}, which would naturally yield a kinetic mixing suppressed by the $A$ and $A'$ gauge couplings,
and thus parametrically identical.
This gives us the more aggressive result that
\begin{equation}
q_\chi = g'\epsilon/e \geq (m_\chi/10\;\rm meV)^4
\label{eq:qDMagg}
\end{equation}
is disallowed by FL, which we also show in Fig.~\ref{fig:MCP}.
While this aggressive constraint can be circumvented if there was a $\epsilon\sim 1$ kinetic mixing in the Lagrangian, or in the (unnatural) case that the unit charge is $q_\chi \ll 1$ where there is no dark photon, the conservative constraint in Eq.~\eqref{eq:qDMcons} is harder to evade.
We emphasize that there are $O(1)$ unknown factors in front of these constraints, and as such they ought to be taken as guidance rather than strict no-go theorems.

We now compare the region covered by the FL arguments with other probes of millicharged particles.
While accelerators are sensitive to new particles up to $\sim$ TeV scale~\cite{Prinz:1998ua,SHIP:2021tpn,LDMX:2018cma}, for the extremely weak charges we are interested there are other more precise laboratory probes.
In particular, PVLAS rules out new light particles with charges large enough to change the vacuum polarization of light~\cite{Ahlers:2006iz,DellaValle:2014xoa}.
We show this limit in Fig.~\ref{fig:MCP}.
That Figure shows that the FL bound improves upon the laboratory limits of PVLAS for $m_\chi \lesssim \mu$eV in the conservative ($\epsilon<1$) case, and for  for $m_\chi \lesssim$ meV for the aggressive ($\epsilon\propto g'$) case,
and continues to strengthen for lower masses.

In addition to laboratory bounds, there are astrophysical arguments, related to cooling of red-giant, horizontal-branch and globular-cluster stars, which can constrain MCPs.
The strongest of these limits is at the $q_\chi\geq 2 \times 10^{-14}$ level~\cite{Davidson:2000hf}.
These limits, however, are indirect, and can be circumvented through model building~\cite{Berlin:2020pey}.
This prompted the proposal of future laboratory experiments to search for MCPs more sensitively in this mass range, and we show the projected reach of the superconductive-cavity experiment from Ref.~\cite{Berlin:2020pey} in Fig.~\ref{fig:MCP}.
We note that our constraints, albeit theoretical in nature, can already disfavor a portion of the parameter space to be probed by these future experiments.
Moreover, our conservative limit can improve upon even the astrophysical constraints for $m_\chi \lesssim 0.1$ neV, or  $m_\chi \lesssim  \mu$eV for our aggressive limit from Eq.~\eqref{eq:qDMagg}.
The constraints we obtain are both phenomenologically relevant and fairly independent of the details of the model\footnote{In \cite{Montero:2019ekk} it was claimed that the constraints on dark photons coming from the FL bound were already superseded by other known constraints. What \cite{Montero:2019ekk} missed is that this is only true within a narrow window at the MeV-GeV scale, and moreover does not cover secluded dark sectors, as we will study next.}.

None of these results require the MCPs to be the cosmological DM.
There are additional limits if the MCPs compose the entirety of the DM, from plasma instabilities~\cite{Lasenby:2020rlf,Cruz:2022otv}, magnetic-field effects~\cite{Kadota:2016tqq,Stebbins:2019xjr}, and coherent effects that de-stabilize DM against annihilations or decays~\cite{Jaeckel:2021xyo} (shown in Fig.~\ref{fig:MCP}).
For MCPs that compose a small fraction of the DM, the effects are more subtle, and include cooling of hydrogen in the early universe~\cite{Munoz:2018pzp}, as well as alter the dispersion relation of radio emission from pulsars~\cite{Caputo:2019tms}.
We mention, in passing, that the extremely light MCPs we consider here would require non-thermal production to be the cosmological DM such as in~\cite{Sikivie:2006ni,Graham:2015rva,Co:2017mop,Agrawal:2018vin,Stebbins:2019xjr}.

\begin{figure}[hbtp!]
\centering
	\includegraphics[width=0.8\textwidth]{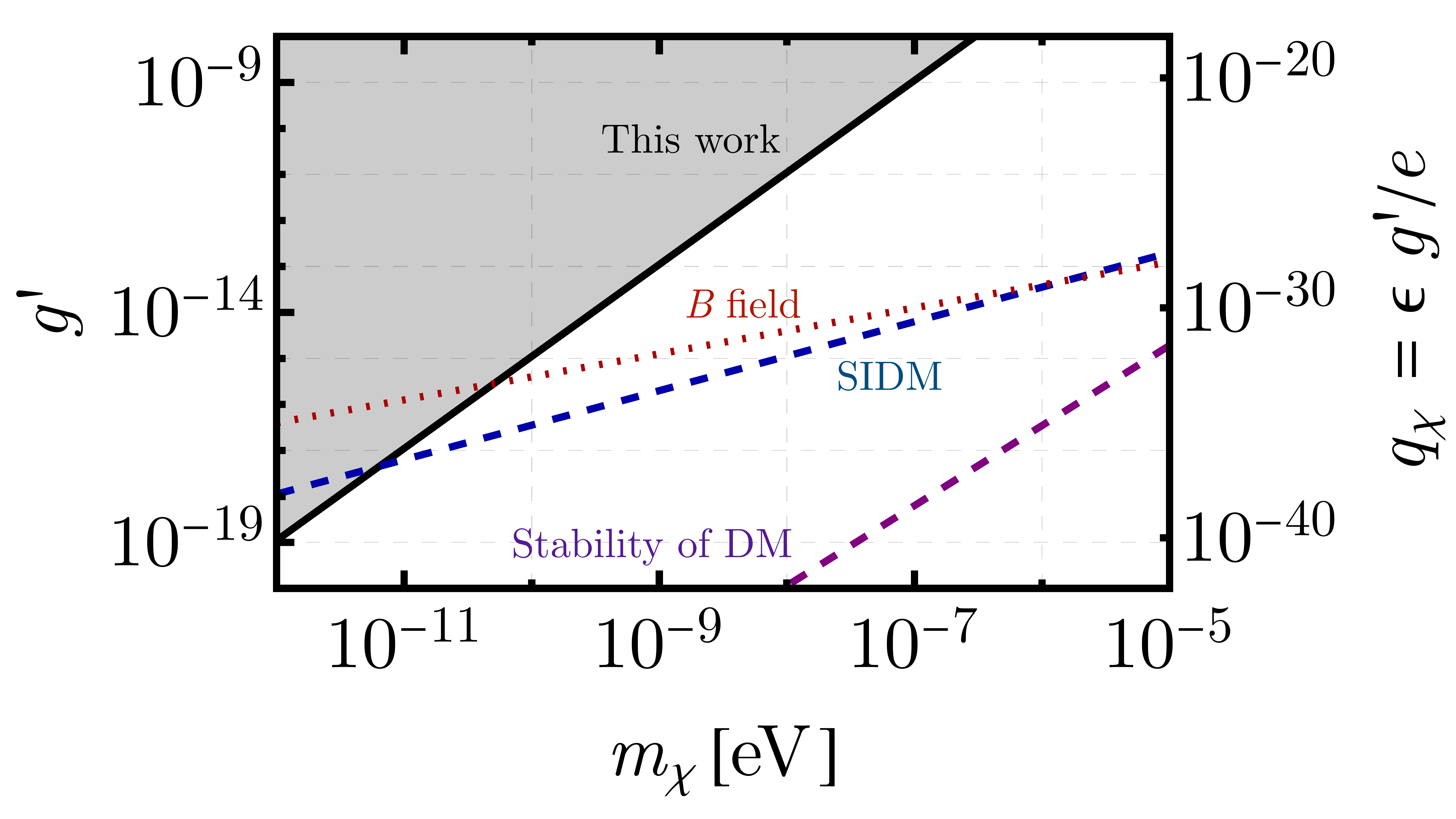}
	\caption{
	Parameter space of darkly charged particles (DCPs), given a coupling $g'$ to a dark photon (assumed massless).
	As in Fig.~\ref{fig:MCP}, the Festina-Lente bound places the black shaded region in the Swampland.
	Different dashed lines show constraints for DCPs if they compose the entirety of the DM, and come from self interactions (blue~\cite{Agrawal:2016quu}), stability arguments (purple~\cite{Jaeckel:2021xyo}), and from magnetic fields (red~\cite{Stebbins:2019xjr}, where we assume an induced millicharge $q_\chi$ given by the standard 1-loop value for the kinetic mixing $\epsilon = e\,g'/(16\pi^2)$).
	}
	\label{fig:MCP_dark}
\end{figure}

\subsection{Case II: Secluded Dark Sectors}

We can additionally apply the FL equation directly on the dark-sector coupling $g'$, regardless of whether the dark $U(1)$ is significantly mixed with our sector.
Such a ``secluded'' dark sector is motivated to be the cosmological DM, as their self interactions can potentially alleviate tensions in structure formation~\cite{Fan:2013yva}.
Assuming a darkly charged particle (DCP), which does not interact with the visible sector other than gravitationally, we find the limits shown in Fig.~\ref{fig:MCP_dark} for different DCP masses $m_\chi$.
These are, to our knowledge, the first time that Swampland limits are applied to dark-sector particles independent on their coupling to our sector.
Additionally, the DM self interactions would affect the shape of galaxies~\cite{Agrawal:2016quu}, as well as alter the famous ``bullet'' cluster collision, which leads to the limit shown in Fig.~\ref{fig:MCP_dark}.
We have additionally shown in Fig.~\ref{fig:MCP_dark} a limit on DM charges from the magnetic field of our galaxy~\cite{Stebbins:2019xjr}, assuming that for any dark charge $g'$ there is a kinetic mixing between the dark and regular photon of size $\epsilon = e g'/(16\pi^2)$, as in the previous subsection.
All these limits, while strong, only apply if a majority of the DM is self-interacting DCPs, whereas our constraint should be present as long as MCPs are in the spectrum of the theory, and the dark photon is lighter than $H_0$.

In the spirit of studying not only the parameter space covered by QG arguments, but also possible model-building to evade them, we note that the arguments underlying the FL bound \eqref{FL} apply only to massless photons (or massive photons but with mass below the Hubble scale $H_0$) \cite{Montero:2019ekk}.
Thus, strictly speaking our constraints may be evaded if the dark photon is made sufficiently massive.
In the next subsection, however, we will use a different Swampland principle to place constraints on dark-photon masses.
While a rigorous argument is lacking, WGC bounds are often true for massive vector bosons as well, so the same may be true for the FL bound.
We do not attempt to generalize these bounds here, but instead mention the massive-dark photon case as a possible loop-hole to our constraints.

\section{St\"{u}ckelberg Dark Photons}
\label{sec:MassiveDarkPhotons}

Massive dark photons have interesting phenomenology (see for example~\cite{Fabbrichesi:2020wbt} for a recent review). Among other things, a massive dark photon can be a viable DM candidate~\cite{Reece:2009un,Graham:2015rva,Agrawal:2018vin,Co:2018lka,Dror:2018pdh,Bastero-Gil:2018uel,Long:2019lwl}, produce observable effects in cosmology (such as altering the 21-cm signal~\cite{Bowman:2018yin,Pospelov:2018kdh}, or heat up the universe~\cite{Kovetz:2018zes}), as well as produce new signatures at colliders~\cite{Pospelov:2008zw,Reece:2009un,BaBar:2014zli,LHCb:2019vmc} or beam dumps~\cite{Bjorken:2009mm,Cesarotti:2022ttv}.
We will now describe Swampland bounds on the mass of the dark photon, first reviewing how it can get a mass.

\subsection{The Mass Term}

The mass $m_{V}$ for a photon $V_\mu$ is described by the corresponding mass term in the Lagrangian,
\begin{equation}
\mathcal{L}\supset -\frac{1}{4} V_{\mu\nu}V^{\mu\nu}- \frac12 g^{\mu\nu} m^2_{V} V_\mu V_{\nu},
\label{orig32}
\end{equation}
where $V_{\mu\nu} = \partial_\mu V_\nu - \partial_\nu V_\mu$. It is often convenient to perform the ``St\"{u}ckelberg trick'' by replacing $V_\mu \rightarrow A_\mu' - \partial_\mu \theta / m_{A'}$ and renaming $m_V \rightarrow m_{A'}$. This allows us to separately describe the longitudinal and transverse polarizations of the massive photon via the equivalent Lagrangian
\begin{equation}
\mathcal{L}'\supset -\frac{1}{4}F'_{\mu\nu}F'^{\mu\nu}
-\frac12 m_{A'}^2g^{\mu\nu}\left(A'_\mu - \frac{\partial_\mu \theta}{m_{A'}} \right) \left(A'_\nu - \frac{\partial_\nu \theta}{m_{A'}} \right),
\label{orig33}
\end{equation}
where now $F_{\mu\nu}'=\partial_\mu A_\nu' - \partial_\nu A_\mu'$, $A'_\mu$ describes the propagation of two degrees of freedom by virtue of the gauge invariance shown below and $\theta$ is a periodic scalar.
The Lagrangian $\mathcal{L}'$ is invariant under the following gauge transformation:
\begin{equation}
A_\mu'\rightarrow A_\mu' + \partial_\mu \lambda, \quad \theta\,\rightarrow\, \theta + m_{A'}\lambda.
\end{equation}
Gauge fixing $\theta=0$ brings us back to unitary gauge and reintroduces a longitudinal polarization into $A_\mu'$, and we recover the original Lagrangian \eqref{orig32}.
What we have described so far is a perfectly valid theory of a free massive photon.
Although completely consistent as a QFT at any energy scale, this theory ought to be UV completed within quantum gravity, and as such the Swampland can shed light on the parameters of this model.
In particular, one must ask about the dynamical origin, in the UV, of the mass term for the photon.
One possibility is that the Lagrangian \eqref{orig33} is describing the low-energy EFT after a charged Higgs\footnote{Note that the field $\Phi$ need not be the usual Higgs, and can be a new ``dark" Higgs that gives mass to the dark sector.} field $\Phi$ picks up a vacuum expectation value (VEV). Using the Lagrangian for a charged field coupled to a massless photon, writing
\begin{equation}
\Phi= h e^{i\varphi},
\end{equation}
and assuming the $U(1)$ theory has gauge coupling $g'$ and the Higgs field picks up a VEV $v$, one recovers at low energies the Lagrangian \eqref{orig33}, with
\begin{equation} m_{A'}= g' v.\end{equation}
In this scenario, there is a massive scalar, the Higgs field, which has a mass naturally of the same order as $A'$. Furthermore this scalar couples to $A'$, since the mass term comes from a coupling
\begin{equation}
  \label{eq:StueckelbergCoupling}
    \mathcal{L} \supset g' m_{A'} h \left(A'_\mu - \frac{1}{m_{A'}} \partial_\mu \theta\right)^2
\end{equation}
where we substituted one $h$ for its VEV.
We will call this scenario, where the theory at energy scales of order $m_{A'}$ becomes that of a massless photon coupled to a charged scalar field, the ``Higgs'' scenario, and will call a mass term arising as above a ``Higgs mass''.  The coupling \eqref{eq:StueckelbergCoupling} is very important for the phenomenology of a Higgs massive dark photon, since it implies the dark photon $A'$ can scatter with the Higgs field $h$ (or at low energies, that the radial mode of $A'$ can be excited), as we will review below.

Within effective field theory, the only other possibility for a massive photon is simply that the St\"{u}ckelberg Lagrangian \eqref{orig32} remains valid even at energy scales beyond $m_{A'}$, all the way to some UV cutoff scale where quantum gravitational (or stringy) effects become strong, possibly even the Planck scale.
We will call this scenario the ``St\"{u}ckelberg mass'' case. This is in contrast to the Higgs case in which~\eqref{orig32} is not valid above $m_{A'}$ where the dynamics of the Higgs also have to be taken into account.

At first sight, the phenomenology of ``St\"{u}ckelberg'' and Higgs massive dark photons seems very different,  owing to the presence or absence of a radial Higgs mode. This also means that different limits can be set on the two scenarios, as outlined in Ref.~\cite{Caputo:2021eaa}.
Broadly speaking, the constraints on the Higgs scenario are stronger due to the presence of a radial mode, which introduces extra interactions that must also be suppressed to be below detection limits.
This is especially important for ultra-light vectors, as otherwise their mixing with the SM is suppressed by the plasma mass of the regular photon.
We show the parameter space of dark photons, given their mass $m_{A'}$ and kinetic mixing $\epsilon$, in Fig.~\ref{fig:SwampStueck}, where it is clear that current astrophysical, laboratory, and DM constraints leave a significant gap for low $m_{A'}$ with weak mixing~\cite{Caputo:2021eaa}.

This region is phenomenologically interesting, as it can possibly account for ultra-light DM, as well as explain the 21-cm excess reported by EDGES~\cite{Bowman:2018yin}.
Here we will study this parameter space under the light of the Swampland, focusing on one guiding principle: that St\"{u}ckelberg photons get their mass by eating an axion. This then leads to two outcomes: the presence of a radial mode (the saxion) and a UV cutoff (set by the tension of axion strings). More details are provided in the following two Subsections.

\subsection{Constraints From Physics of the Radial Mode}
\label{sec:radialmode}

A key recent insight is that ``pure St\"{u}ckelberg masses'' are in the Swampland.
In more detail, the argument put forth in Ref.~\cite{Reece:2018zvv} is that whenever one finds an axion $a$ in a consistent quantum theory of gravity, it is always accompanied by a corresponding ``radial mode''. This radial mode is often called a saxion, because together with the axion they constitute the two real scalars of an $\mathcal{N}=1$ chiral multiplet, but the statement is supposed to hold even when SUSY is broken.  The claim was first introduced in one of the original Swampland papers~\cite{Ooguri:2006in}, where it was mapped to the geometrical property that there is no closed curve of minimum length in the moduli space. This statement has significant phenomenological implications; for instance, under some mild assumptions, it allows one to conclude that the Standard Model photon must be exactly massless \cite{Reece:2018zvv}.

This radial mode, which we will call $\sigma$, has couplings very similar to those of the Higgs. In particular, a coupling like Eq.~\eqref{eq:StueckelbergCoupling} but with $h \rightarrow \sigma$, always exists as a consequence of the fact that the radial mode allows for the shrinking of the closed curve in scalar space parametrized by the axion. The basic conclusion is thus that the existence of the radial mode---and its coupling---render the St\"{u}ckelberg case somewhat similar to the Higgs case. This can give us an extra handle to disfavor new regions of parameter space, as we can import constraints on dark photons from the Higgs to the St\"{u}ckelberg scenarios. There is an important caveat to this reasoning that we outline below.

An important feature of the Higgs scenario is that there is a precise prediction for the mass of the Higgs mode and, barring tuning, it is of the same order as the mass of the vector boson. In known stringy examples, this is also true of the mass of the radial mode. Phrased in terms of the dark-photon mass, this upper limit on the radial-mode mass is:
\begin{align}
    m_\sigma \leq 4\pi m_{A'}/g'
    \label{eq:sigmamass}
\end{align}
where $m_{A'}$ is the mass of the dark photon and $g'$ is the dark gauge coupling. One can give a heuristic argument for \eqref{eq:sigmamass} roughly as follows. The dual field to the axion couples to strings, which must satisfy a version of the WGC for axions \cite{Hebecker:2017wsu}. This sets a cutoff for the effective field theory,
\begin{equation}\Lambda_{\text{UV}}\sim \sqrt{f M_{\rm Pl}}\sim \sqrt{\frac{m_{A'}M_\mathrm{Pl}}{g'}}.\label{eq:cutoffstr}\end{equation}
According to the Swampland Distance Conjecture \cite{Ooguri:2006in}, we expect the effective field theory to be valid for saxion field displacements $\Delta\sigma\sim\mathcal{O}(M_{\rm Pl})$. In such a variation, the potential energy increases by
\begin{equation}\Delta V\sim m_\sigma^2 \Delta \sigma^2= m_\sigma^2\,M_{\rm Pl}^2.\end{equation}
Imposing that this variation is describable within the EFT leads to $\Delta V\leq \Lambda_{\text{UV}}^4$, which, when rearranged, leads to \eqref{eq:sigmamass}.
Taking \eqref{eq:sigmamass} into account, the phenomenology of Higgs and St\"{u}ckelberg scenarios is similar due to Swampland constraints. For instance, in the Higgs scenario, there is a coupling
\begin{equation}\frac{m}{g'}\, h \, (\partial \theta)^2\label{e334}\end{equation}
which is relevant for stellar-cooling constraints. We now show that a similar coupling is present in the Stuckelberg case. The universal asymptotic structure for an $\mathcal{N}=1$ kinetic term for the saxion-axion system is
\begin{equation}\mathcal{L}\supset\frac{M_*^2}{4s^2}\left(d\sigma^2+d\theta^2\right). \label{w2}\end{equation}
This structure is not only present in all 4d $\mathcal{N}=1$ limits in known string theory compactifications; it is also true in non-supersymmetric setups like the $O(16)\times O(16)$ string \cite{Alvarez-Gaume:1986ghj}, and it is indeed part of the motivation behind the radial-mode conjecture of \cite{Reece:2018zvv}. Working around a particular expectation value $s_0=\langle s\rangle$ for the saxion, we can introduce the physical axion decay constant
\begin{equation} f\equiv \frac{M_*^2}{2s_0^2}\end{equation}
which allows us to rewrite \eqref{w2} as
\begin{equation}\frac{f^2}{2}\left(1-2\sqrt{2}\frac{f}{M_*}\sigma+\ldots\right)(d\sigma^2+d\theta^2). \label{w3}\end{equation}
Canonically normalizing the field $\sigma$ as $\hat{\sigma}=f\sigma$, we obtain a coupling
\begin{equation}
\mathcal{L}\supset \sqrt{2}\frac{\hat\sigma}{M_*}(f\, \partial \theta)^2
=\sqrt{2}\frac{f}{M_*}f \hat\sigma ( \partial \theta)^2
= \left(\sqrt{2}\frac{m}{g'\, M_*}\right)\frac{m}{g'} \hat\sigma ( \partial \theta)^2\end{equation}

\noindent in the Lagrangian. This has  the same structure as the coupling \eqref{e334}, with an additional suppression by a factor of
\begin{equation}\alpha\equiv \sqrt{2}\frac{m}{g'\, M_*}.
\label{eq:alphadef}
\end{equation}
This is the caveat we previously mentioned: the two cases are very similar up to the parametrics of the coupling between the radial mode and the massive photon.

\begin{figure}[hbtp!]
    \centering
	\includegraphics[width=0.36\textwidth]{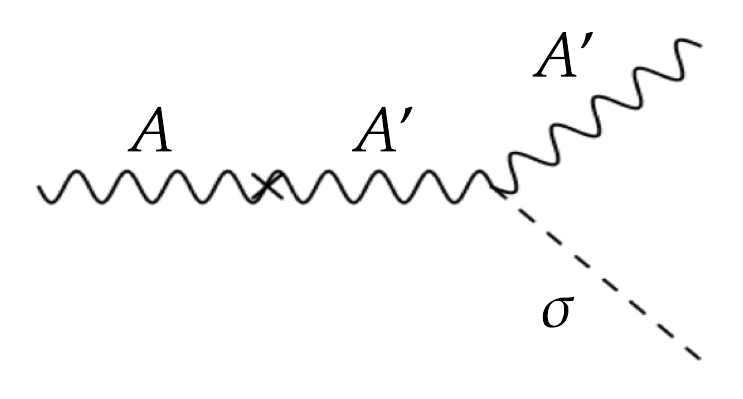}
	\caption{The `Higgs-strahlung' process for the St\"{u}ckelberg photon. In this case, the emitted scalar particle is $\sigma$, the radial mode required by the radial mode conjecture (see Section~\ref{sec:radialmode}). The amplitude for this process is proportional to the product $\epsilon \alpha g'$
	where the first factor comes from the kinetic mixing and the second from the vertex (see Eq.~\eqref{eq:alphadef} for the definition of $\alpha$).}
	\label{fig:Higgsstrahlung}
\end{figure}

The fact that the light radial mode (Eq.~\eqref{eq:sigmamass}) couples to the St\"{u}ckelberg photon (as in Eq.~\eqref{eq:StueckelbergCoupling} with $h\rightarrow \alpha \sigma$) means that we have similar
interactions in the St\"{u}ckelberg case as in the Higgs case. To estimate the constraints on massive St\"{u}ckelberg photons, we consider the implications of these couplings for stellar-cooling effects. In particular, the Higgs-strahlung process in Fig.~\ref{fig:Higgsstrahlung} that is known to dominate the production of light dark Higgses and photons in stellar plasmas~\cite{Pospelov:2008jk,Batell:2009yf,An:2013yfc} also exists for the St\"{u}ckelberg photon (Fig.~\ref{fig:Higgsstrahlung}).

It is immediately clear from inspecting the cross term in~\eqref{eq:StueckelbergCoupling} that the amplitude for the process $A\rightarrow A'_T \rightarrow h + \theta$ is proportional to the product $\epsilon \alpha g' = \sqrt{2}\epsilon m/M_*$. This process is important as long as there is enough energy to allow for the production of the final state. In practice, this means that the masses of $A'$ and $\sigma$ should be lower than the plasma mass of
the photon at the Sun, which is $\mathcal{O}(100\;\mathrm{eV})$. For these plasma decay processes not to significantly alter stellar evolution, the amplitude of interactions such as those shown in Fig.~\ref{fig:Higgsstrahlung} should be small. This criterion has been used to constrain the importance of similar processes in the case of Higgsed massive photons~\cite{An:2013yfc} and millicharged particles~\cite{Davidson:1991si,Davidson:2000hf}.
In our case, the limits from~\cite{Davidson:2000hf} imply:
\begin{align}
    \sqrt{2} \frac{\epsilon m}{M_*} < 10^{-14}.
    \label{eq:mixinglimit}
\end{align}
This is shown in Figure~\ref{fig:SwampStueck} for two representative values of $M_*$. Because of the smallness of $m$, the only relevant constraint is when $M_*$ is also small. In string models, $M_*$ is typically the string scale and experimental constraints would prevent us from setting it too low. However, if there are models where $M_*$ is effectively replaced by a low scale, then our bound above can become more constraining.

We finally note that the bound \eqref{eq:sigmamass} can be avoided by making the Higgs $h$ (or radial mode $\sigma$ for the St\"{u}ckelberg case) heavy enough, so it is not produced in stellar environments, at the cost of fine tuning.
For the Higgs case, one can attempt to set
\begin{equation}
m_h \gg m_{A'}\quad\Rightarrow\quad \sqrt{\lambda} \gg g', \label{eq:pothi}
\end{equation}
where $\lambda$ is the Higgs quartic coupling, and $g'$ the dark gauge coupling as before. Perturbative unitarity places an upper limit on $\lambda \lesssim 8\pi^2$. The above hierarchy then has to be arranged by choosing small $g'$. For example, to have dark photons in the $m_{A'} \sim 10^{-14}$ eV mass range, but $m_h \gtrsim $ keV (so as to avoid the stellar constraints), one would need ${g'}/\sqrt{\lambda} \lesssim 10^{-17}$, i.e., extremely feebly interacting dark charges. The question of whether potentials with a large hierarchy such as \eqref{eq:pothi} are in the Landscape or in the Swampland is very interesting (indeed, it is tantamount to the Electroweak hierarchy problem), but beyond the scope of this paper.
The St\"{u}ckelberg scenario follows identically, as the bound for $m_\sigma$ in Eq.~\eqref{eq:sigmamass} (from Ref.~\cite{Reece:2018zvv}) comes from a similar argument.
In this case, we can phrase the tuning in terms of requiring a kinetic mixing several orders of magnitude larger than implied by the formula $\epsilon\sim e g'/(16\pi^2)$.

\subsection{Constraints from the Axion String Species Scale}

There is another constraint that follows from the arguments in \cite{Reece:2018zvv}, which applies only in the St\"{u}ckelberg case. In a St\"{u}ckelberg theory, the axion $\theta$ is ``fundamental'', in the sense that it is not replaced by any other degree of freedom before the EFT breaks down. Consequently, one can consider metastable axion strings to which one may apply the corresponding version of the WGC. Doing this, one can put a bound on the string tension, and a corresponding string scale, which sets an upper bound $\Lambda_\mathrm{UV}$ for the cutoff of the local effective field theory as\footnote{This cutoff assumes a ``strong form'' of the WGC for strings, which amounts to assuming that the bound is satisfied by an object of charge one. Relaxing this assumption may relax the cutoff. If the object that satisfies the WGC has charge $n$, the upper bound described in the main text relaxes by a factor of $\sqrt{n}$. However, in all stringy examples known to date, $n\sim \mathcal{O}(1)$. Proving that $n$ is always small remains an important open question in the Swampland program.}
\begin{equation}
\Lambda_\mathrm{UV} < \sqrt{m_{A'} M_\mathrm{Pl} / g'}.
\label{eq:UVcutoff}
\end{equation}
We remind the reader that the physical interpretation of the cutoff \eqref{eq:UVcutoff}, which also appeared in \eqref{eq:cutoffstr}, is that it corresponds to the energy scale of the strings that couple magnetically to the axion. At this scale, therefore, a fundamental string becomes light, and the local effective field theory description breaks down (indeed, $\Lambda_\mathrm{UV}$ corresponds to the ``species scale'' in the case of a perturbative string limit). Since quantum field theory remains applicable at energy scales probed by the LHC, one must require this cutoff to be above  $\approx$ 10 TeV.
This gives:
\begin{align*}
  g ' \leq \frac{m_{A'}}{40\;\mathrm{meV}}.
\end{align*}
Assuming that the magnitude of kinetic mixing is given by Eq.~\ref{eq:oneLoopMixing}, we obtain a bound:
\begin{align}
  \epsilon \approx \frac{e g'}{16 \pi^2} \lesssim  \frac{m_{A'}}{20\;\mathrm{eV}}.
  \label{eq:limitUV}
\end{align}
This is also shown in Fig.~\ref{fig:SwampStueck}.

\subsection{Phenomenological Implications}

\begin{figure}[hbtp!]
    \centering
	\includegraphics[width=0.85\textwidth]{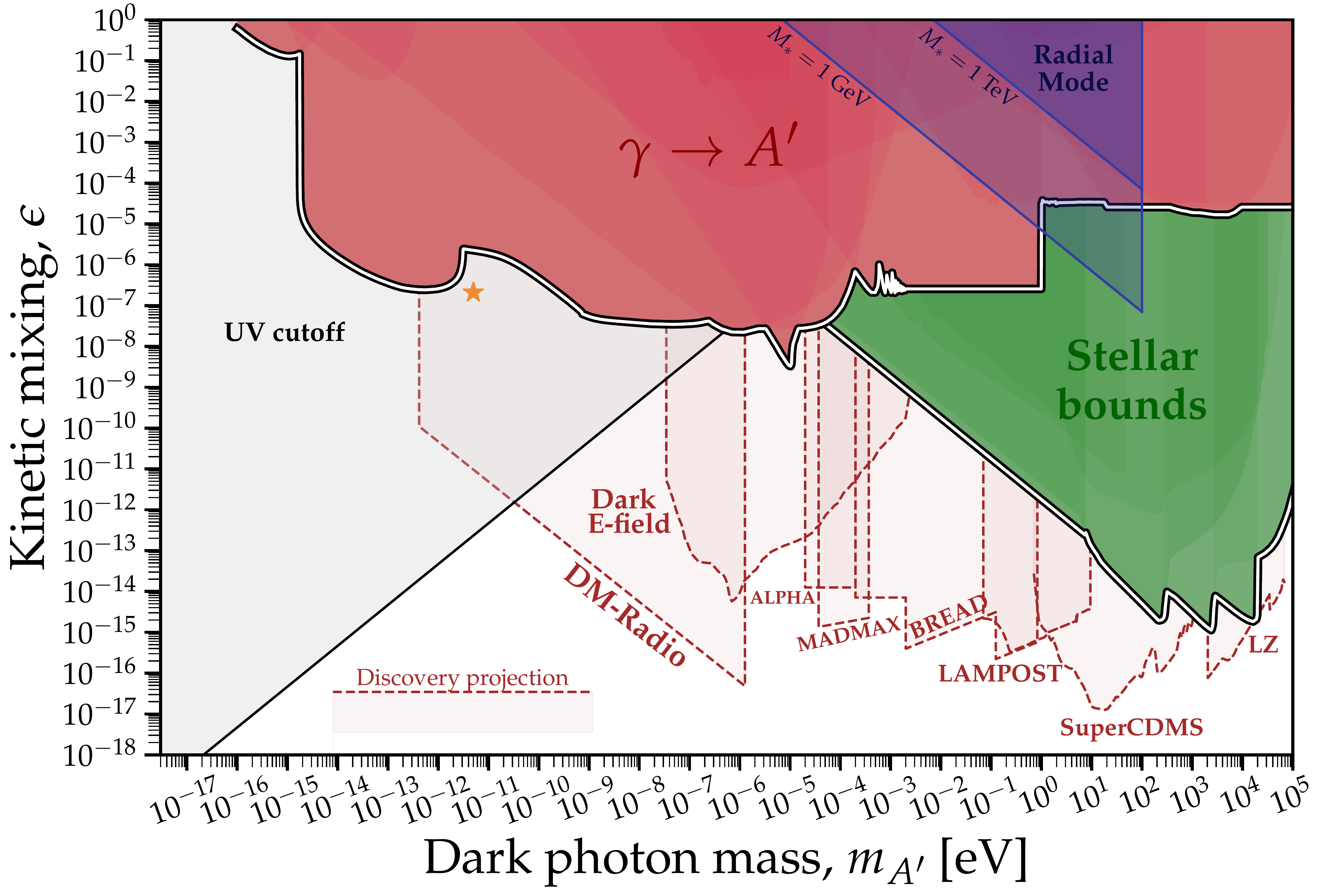}
	\caption{
	Parameter space of a St\"{u}ckelberg dark photon. The red region is constrained by experiments that measure photon-to-dark photon transitions,
	such as COBE/FIRAS~\cite{Fixsen:1996nj} and light-shining-through-walls experiments~\cite{Ehret:2010mh,Inada:2013tx,Povey:2010hs,Parker:2013fxa,ADMX:2010ubl,Betz:2013dza}.
	The green region is constrained by bounds on stellar cooling from the Sun, Horizontal Branch stars and Red Giants~\cite{Redondo:2013lna}.
	The gray shaded region is ruled due to the UV cutoff from Eq.~\eqref{eq:limitUV}. The blue region is ruled out following the existence of the radial mode from Eq.~\eqref{eq:mixinglimit}, for which we show two values of $M_*$.  The orange star shows the model of~\cite{Pospelov:2018kdh} which is an attempt to explain the 21 cm EDGES anomaly and is disfavored by our constraints.
    We note that there are further constraints from the Solar basin around eV masses~\cite{Lasenby:2020goo}, as well as from superradiance for lower masses~\cite{Cardoso:2018tly}.
	}
	\label{fig:SwampStueck}
\end{figure}

The Swampland-disfavored regions that we show in Fig.~\ref{fig:SwampStueck} have important phenomenological implications, which we now explore.

First, there has been considerable attention on the low-frequency tail of the cosmic microwave background (CMB) as a possible avenue for new physics, motivated by the 21-cm detection during cosmic dawn by the EDGES collaboration~\cite{Bowman:2018yin}.
A class of models that attempt to explain the signal introduce a dark photon that oscillates to the SM photon thereby increasing the number of CMB photons in the Rayleigh-Jeans tail of the distribution, e.g.~\cite{Pospelov:2018kdh,Garcia:2020qrp,Caputo:2020bdy}. The increased photon number acts as an extra radio background, and thus deepens the 21-cm absorption trough, as claimed by EDGES. The dark-photon parameters proposed in Ref.~\cite{Pospelov:2018kdh} are indicated by the yellow star in Fig.~\ref{fig:SwampStueck}.
In order for this model to avoid our constraint from Eq.~\eqref{eq:limitUV} one would need $g' \lesssim 10^{-10}$ to give rise to $\Lambda_{\rm UV} >$ 10 TeV.
This demands a fair amount of fine tuning, or a new mechanism to give rise to small kinetic mixing other than the usual 1-loop term from Ref.~\cite{Holdom:1985ag} (as $\epsilon \propto g'$ would be far too large with $\mathcal{O}(1)$ couplings).

Second, the dark-photon portal is one of the most popular avenues to a renormalizable theory of dark matter that interacts with our sector~\cite{Chu:2011be}.
In particular, freeze-in DM~\cite{Hall:2009bx,Bernal:2017kxu}, the case in which the DM is slowly produced from interactions with our sector over cosmic history but never thermalizes, is tantalizingly close to the reach of direct-detection experiments.
In this scenario the DM coupling to our sector is through a very light mediator, and it requires tiny couplings ($\sim 10^{-12}$, e.g.~\cite{Dvorkin:2019zdi}).
Our results imply that freeze-in through a kinetically mixed dark photon is disfavored for masses $m_{A'} < 10^{-10}$ eV (assuming the standard $\epsilon \propto g'$ mixing), given the $\epsilon \times g'< 10^{-14}$ requirement for freeze-in\footnote{Note that for freeze-in through the dark photon (rather than the SM plasma), which would only depend on $g'$ and not $\epsilon$, this particle has to be cosmologically populated, and this is constrained by measurements of the relativistic number $N_{\rm eff}$ of degrees of freedom at BBN for $m_{A'}\lesssim$ MeV.}.

Finally, ultralight bosons are an attractive dark-matter candidate, and dark photons in particular can produce the correct DM abundance for masses as low as $m_{A'}\sim 10^{-20}$ eV~\cite{Agrawal:2018vin,Co:2018lka}.
A plethora of experimental efforts have been developed to test $A'$ DM, and we show the reach of some of those ``direct-detection'' experiments in Fig.~\ref{fig:SwampStueck}.
Our work shows that part of the parameter space targeted by these experiments for low $m_{A'}$ is in the Swampland.
Therefore, a detection of a dark photon within this region is not expected, and it would test our knowledge of quantum gravity.

\begin{section}{Non-Abelian Dark Matter}
\label{sec:NAmodels}
In this Section we constrain Dark Matter models that make use of non-Abelian gauge fields. The idea is simple  and relies on the Festina Lente bound described above. As noted in~\cite{Montero:2019ekk} and reviewed in Section \ref{sec:FLbound}, the existence of a lower bound on the mass of charged particles in (quasi-)de Sitter space means that unconfined and un-Higgsed non-Abelian gauge fields are forbidden since their gluons are massless charged particles. These gluons catalyze the decay of Nariai black holes leading to pathological spacetimes with naked singularities. While a few applications have been pointed out in~\cite{Montero:2019ekk,Montero:2021otb}, we take this opportunity to apply this bound to a specific model and comment on potential general lessons that we can learn.
We attempt to describe the model being discussed in a self-contained manner to bring out the features related to the non-Abelian gauge fields and show their appeal for model-building. In this section, we will only discuss models that describe dark-matter physics, leaving comments and related ideas about inflationary and dark-energy models to Appendix~\ref{app:FLandInflation} and Appendix~\ref{app:CCrelaxion}, respectively.

The specific non-Abelian Dark Matter (NADM) model we study was proposed in~\cite{Buen-Abad:2015ova}.
The DM candidate in the NADM model is a WIMP (Weakly Interacting Massive Particle) that is thermally produced in the early universe. In addition, it transforms in the fundamental representation of a dark $SU(N)_d$ gauge group. The gluons of the dark non-Abelian symmetry are weakly coupled today and constitute a dark radiation (DR) component that interacts with the DM. The lack of observation of DM self-interactions places an upper bound on the dark gauge coupling which gives $g_d < 10^{-3}$. The presence of the interacting DM-DR system as well as the multiplicity of the DM lead to the distinguishing features of the NADM model. For instance, its effect on $N_{\rm eff}$ presents an opportunity to relieve the Hubble tension (see e.g.~\cite{Verde:2019ivm}) by altering the CMB prediction of the Hubble expansion rate $H_0$. In addition, the DM multiplicity decreases the cross-section relevant for indirect-detection experiments since the DM color degrees of freedom must match for successful annihilation (see also~\cite{Chen:2009ab} for example).
By contrast, the cross-section seen in collider experiments will be enhanced since colliders can produce any of the $N$ DM particles in the final state. Finally, DM-DR interactions can potentially play a role in the resolution of the $\sigma_8$ tension \cite{Battye:2014qga} since they can lead to a smooth suppression of the matter power spectrum rather than a sharp cut-off at small scales.

Since the dark gluons are weakly coupled at galactic scales, the confinement scale $\Lambda_{\rm conf.} \propto e^{-1/g_d^2}$ is much below Hubble, and the model is incompatible with FL. In particular, in this theory, one can start with a Nariai black hole that has a charge along a Cartan direction of the non-Abelian gauge group and the massless gluons would immediately screen this charge causing the black hole solution to leave the extremality region as shown in Fig.~\ref{fig:FLDiagram}.
We emphasize that this conclusion does not require a relic abundance of non-Abelian dark radiation, as is assumed in~\cite{Buen-Abad:2015ova}, and is therefore phenomenologically stronger than an experimental exclusion since the latter can potentially be avoided by reducing the abundance of non-Abelian gauge particles.

We briefly mention that we do not rule out the features of this model but only this particular realization. For example, one can have a bath of interacting relativistic components to serve as interacting radiation in lieu of the gluon bath and the FL bound would not necessarily apply in such settings. Other features may be more difficult to reproduce. An example is the DM multiplicity. This leads to correlated enhancement/suppression of the cross-section seen in various DM experiments and this signal might be difficult to come by without the presence of a symmetry. In the NADM model, the $SU(N)_d$ gauge group ensures the spectrum has this symmetry. Instead of a gauge symmetry, one may attempt to use a global symmetry which is broken at a high scale (cf.~\cite{Banks:2010zn}). The absence of gauge bosons, however, will change the interaction pattern between DM particles leading to very different phenomenology that deserves an independent study.

\end{section}

\section{Conclusions}
\label{sec:conclusions}

New dark sectors are ubiquitous in extensions of the Standard Model of particle physics, and in fact are necessary to explain the existence of dark matter and dark energy in our universe.
These dark sectors, for instance composed of darkly charged particles or massive dark photons, are well-motivated dark-matter candidates, as well targets for new-physics searches.
As such, any new insight on their particle content can become invaluable.

Dark sectors can leave distinct signatures in both particle-physics experiments and cosmological observables.
This has led to a very active research field aimed at covering the vast range of possible models.
In this note we have added to this rich experimental landscape by studying which parts of their parameter space contradict our current understanding of Quantum Gravity (QG).
For that, we have used recent advances in the Swampland program.

While the usual point of view is that QG is far beyond experimental reach, given the remoteness of the Planck scale, the Swampland program uses insights from unitarity and properties of black holes to constrain them.
Using these principles, it is possible to place interesting constraints in low-energy effective field theories, which can then be checked against a plethora of String Theory constructions (so that String Theory here acts as a ``laboratory'' to check proposed Swampland constraints), and applied to phenomenologically interesting models.

In this paper, we have done exactly that, using the principles of~\cite{Reece:2018zvv} and the Festina Lente (FL) bound of \cite{Montero:2019ekk} to constrain models of dark matter, as well as dark energy and inflation. Some of these models (most notably non-Abelian dark matter) are in the Swampland according to these principles.
As a consequence, the phenomenology they predict, if observed, must be due to different physics.

We have also been able to significantly constrain the parameter space of charged dark matter, both in the case of secluded hidden sectors as well as ``millicharged'' DM, where the FL bound covers previously allowed regions of parameter space for low DM masses $m_\chi \lesssim \mu$eV.
Moreover, we have used the existence of the radial mode to place constraints on the St\"{u}ckelberg mass and kinetic mixing of dark photons.
This, together with previous constraints on Higgsed dark photons~\cite{Ahlers:2008qc,An:2013yfc,An:2013yua}, disfavors the entire region of $m_{A'}\lesssim 20 \,\epsilon$ eV (under the assumption of a standard $\epsilon \propto g'$ kinetic mixing).
These bounds on the dark sector are a novel application of Swampland principles, and can shed light on the long-standing puzzle of the nature of DM.

Given the strength of the FL bound, it is natural to expect that the results in this paper can be extended, resulting in more general and far-reaching constraints. One very interesting avenue is studying the implication of the FL bound on inflation, as we have only scratched the surface in this work. See also \cite{Lee:2021cor,Guidetti:2022xct} for recent work on this direction.

As we have shown here,  ``theoretical'' probes from the Swampland are highly complementary to the observational program already underway to detect the dark sector of our Universe.
It is interesting that while both quantum gravity and the dark sector of our cosmos are open questions in Physics, we can make progress by combining our limited understanding of both these areas.
Though the Swampland is a very active topic on the formal side, there has been so far little exploration of the phenomenological implications of existing Swampland bounds,
as evident by the scarcity of Swampland literature on such an important topic as dark matter phenomenology (see, however, \cite{Shiu:2013wxa,McDonough:2021pdg,Heidenreich:2016aqi,Heidenreich:2017sim}, and especially \cite{Noumi:2022zht}, which bounds the parameter space of dark photons from positivity arguments).
Our hope is that this paper will entice more phenomenology and astrophysics experts to uncover the consequences of Quantum Gravity for our universe.

\vspace{0.5cm}
\textbf{Acknowledgements} We thank Peter Adshead, Cari Cesarotti, Luis Iba\~{n}ez,Rashmish Mishra, Alberto Nicolis, Matthew Reece, Martin Schmaltz, Washington Taylor,  Andrew Tolley, Cumrun Vafa, and  Marcus Aurelius for many interesting discussions and comments in the manuscript. The work of MM is supported by a grant from the Simons Foundation (602883, CV) and by the NSF grant PHY-2013858.
JBM is supported by a Clay Fellowship at the Smithsonian Astrophysical Observatory. The work of GO is supported in part by a grant from the Simons Foundation
(602883, CV).

\bibliography{references}

\providecommand{\href}[2]{#2}\begingroup\raggedright\begin{thebibliography}{100}

\bibitem{Muong-2:2021ojo}
{\scshape Muon g-2} collaboration, \emph{{Measurement of the Positive Muon
  Anomalous Magnetic Moment to 0.46 ppm}},
  \href{https://doi.org/10.1103/PhysRevLett.126.141801}{\emph{Phys. Rev. Lett.}
  {\bfseries 126} (2021) 141801}
  [\href{https://arxiv.org/abs/2104.03281}{{\ttfamily 2104.03281}}].

\bibitem{LHCb:2021trn}
{\scshape LHCb} collaboration, \emph{{Test of lepton universality in
  beauty-quark decays}},
  \href{https://doi.org/10.1038/s41567-021-01478-8}{\emph{Nature Phys.}
  {\bfseries 18} (2022) 277}
  [\href{https://arxiv.org/abs/2103.11769}{{\ttfamily 2103.11769}}].

\bibitem{cdf2022high}
C.~Collaboration†‡, T.~Aaltonen, S.~Amerio, D.~Amidei, A.~Anastassov,
  A.~Annovi et~al., \emph{High-precision measurement of the w boson mass with
  the cdf ii detector}, {\emph{Science} {\bfseries 376} (2022) 170}.

\bibitem{Hui:2016ltb}
L.~Hui, J.P.~Ostriker, S.~Tremaine and E.~Witten, \emph{{Ultralight scalars as
  cosmological dark matter}},
  \href{https://doi.org/10.1103/PhysRevD.95.043541}{\emph{Phys. Rev. D}
  {\bfseries 95} (2017) 043541}
  [\href{https://arxiv.org/abs/1610.08297}{{\ttfamily 1610.08297}}].

\bibitem{Carr:2021bzv}
B.~Carr and F.~Kuhnel, \emph{{Primordial black holes as dark matter
  candidates}},
  \href{https://doi.org/10.21468/SciPostPhysLectNotes.48}{\emph{SciPost Phys.
  Lect. Notes} {\bfseries 48} (2022) 1}
  [\href{https://arxiv.org/abs/2110.02821}{{\ttfamily 2110.02821}}].

\bibitem{Roszkowski:2017nbc}
L.~Roszkowski, E.M.~Sessolo and S.~Trojanowski, \emph{{WIMP dark matter
  candidates and searches\textemdash{}current status and future prospects}},
  \href{https://doi.org/10.1088/1361-6633/aab913}{\emph{Rept. Prog. Phys.}
  {\bfseries 81} (2018) 066201}
  [\href{https://arxiv.org/abs/1707.06277}{{\ttfamily 1707.06277}}].

\bibitem{XENON:2018voc}
{\scshape XENON} collaboration, \emph{{Dark Matter Search Results from a One
  Ton-Year Exposure of XENON1T}},
  \href{https://doi.org/10.1103/PhysRevLett.121.111302}{\emph{Phys. Rev. Lett.}
  {\bfseries 121} (2018) 111302}
  [\href{https://arxiv.org/abs/1805.12562}{{\ttfamily 1805.12562}}].

\bibitem{LUX:2015abn}
{\scshape LUX} collaboration, \emph{{Improved Limits on Scattering of Weakly
  Interacting Massive Particles from Reanalysis of 2013 LUX Data}},
  \href{https://doi.org/10.1103/PhysRevLett.116.161301}{\emph{Phys. Rev. Lett.}
  {\bfseries 116} (2016) 161301}
  [\href{https://arxiv.org/abs/1512.03506}{{\ttfamily 1512.03506}}].

\bibitem{Raffelt:1996wa}
G.G.~Raffelt, \emph{{Stars as laboratories for fundamental physics}: {The
  astrophysics of neutrinos, axions, and other weakly interacting particles}},
  Theoretical Astrophysics, University of Chicago Press (5, 1996).

\bibitem{Alvarez:2014vva}
M.~Alvarez et~al., \emph{{Testing Inflation with Large Scale Structure:
  Connecting Hopes with Reality}},
  \href{https://arxiv.org/abs/1412.4671}{{\ttfamily 1412.4671}}.

\bibitem{CMB-S4:2016ple}
{\scshape CMB-S4} collaboration, \emph{{CMB-S4 Science Book, First Edition}},
  \href{https://arxiv.org/abs/1610.02743}{{\ttfamily 1610.02743}}.

\bibitem{Vafa:2005ui}
C.~Vafa, \emph{{The String landscape and the swampland}},
  \href{https://arxiv.org/abs/hep-th/0509212}{{\ttfamily hep-th/0509212}}.

\bibitem{Brennan:2017rbf}
T.D.~Brennan, F.~Carta and C.~Vafa, \emph{{The String Landscape, the Swampland,
  and the Missing Corner}},
  \href{https://doi.org/10.22323/1.305.0015}{\emph{PoS} {\bfseries TASI2017}
  (2017) 015} [\href{https://arxiv.org/abs/1711.00864}{{\ttfamily
  1711.00864}}].

\bibitem{Palti:2019pca}
E.~Palti, \emph{{The Swampland: Introduction and Review}},
  \href{https://doi.org/10.1002/prop.201900037}{\emph{Fortsch. Phys.}
  {\bfseries 67} (2019) 1900037}
  [\href{https://arxiv.org/abs/1903.06239}{{\ttfamily 1903.06239}}].

\bibitem{vanBeest:2021lhn}
M.~van Beest, J.~Calder\'on-Infante, D.~Mirfendereski and I.~Valenzuela,
  \emph{{Lectures on the Swampland Program in String Compactifications}},
  \href{https://arxiv.org/abs/2102.01111}{{\ttfamily 2102.01111}}.

\bibitem{Montero:2019ekk}
M.~Montero, T.~Van~Riet and G.~Venken, \emph{{Festina Lente: EFT Constraints
  from Charged Black Hole Evaporation in de Sitter}},
  \href{https://doi.org/10.1007/JHEP01(2020)039}{\emph{JHEP} {\bfseries 01}
  (2020) 039} [\href{https://arxiv.org/abs/1910.01648}{{\ttfamily
  1910.01648}}].

\bibitem{ArkaniHamed:2006dz}
N.~Arkani-Hamed, L.~Motl, A.~Nicolis and C.~Vafa, \emph{{The String landscape,
  black holes and gravity as the weakest force}},
  \href{https://doi.org/10.1088/1126-6708/2007/06/060}{\emph{JHEP} {\bfseries
  06} (2007) 060} [\href{https://arxiv.org/abs/hep-th/0601001}{{\ttfamily
  hep-th/0601001}}].

\bibitem{Reece:2018zvv}
M.~Reece, \emph{{Photon Masses in the Landscape and the Swampland}},
  \href{https://doi.org/10.1007/JHEP07(2019)181}{\emph{JHEP} {\bfseries 07}
  (2019) 181} [\href{https://arxiv.org/abs/1808.09966}{{\ttfamily
  1808.09966}}].

\bibitem{Obied:2021zjc}
G.~Obied and A.~Parikh, \emph{{A Tale of Two $U(1)$'s: Kinetic Mixing from
  Lattice WGC States}},  \href{https://arxiv.org/abs/2109.07913}{{\ttfamily
  2109.07913}}.

\bibitem{Cheung:2018cwt}
C.~Cheung, J.~Liu and G.N.~Remmen, \emph{{Proof of the Weak Gravity Conjecture
  from Black Hole Entropy}},
  \href{https://doi.org/10.1007/JHEP10(2018)004}{\emph{JHEP} {\bfseries 10}
  (2018) 004} [\href{https://arxiv.org/abs/1801.08546}{{\ttfamily
  1801.08546}}].

\bibitem{Hamada:2018dde}
Y.~Hamada, T.~Noumi and G.~Shiu, \emph{{Weak Gravity Conjecture from Unitarity
  and Causality}},
  \href{https://doi.org/10.1103/PhysRevLett.123.051601}{\emph{Phys. Rev. Lett.}
  {\bfseries 123} (2019) 051601}
  [\href{https://arxiv.org/abs/1810.03637}{{\ttfamily 1810.03637}}].

\bibitem{Arkani-Hamed:2021ajd}
N.~Arkani-Hamed, Y.-t.~Huang, J.-Y.~Liu and G.N.~Remmen, \emph{{Causality,
  unitarity, and the weak gravity conjecture}},
  \href{https://doi.org/10.1007/JHEP03(2022)083}{\emph{JHEP} {\bfseries 03}
  (2022) 083} [\href{https://arxiv.org/abs/2109.13937}{{\ttfamily
  2109.13937}}].

\bibitem{Grimm:2019ixq}
T.W.~Grimm, C.~Li and I.~Valenzuela, \emph{{Asymptotic Flux Compactifications
  and the Swampland}},
  \href{https://doi.org/10.1007/JHEP06(2020)009}{\emph{JHEP} {\bfseries 06}
  (2020) 009} [\href{https://arxiv.org/abs/1910.09549}{{\ttfamily
  1910.09549}}].

\bibitem{Harlow:2022gzl}
D.~Harlow, B.~Heidenreich, M.~Reece and T.~Rudelius, \emph{{The Weak Gravity
  Conjecture: A Review}},  \href{https://arxiv.org/abs/2201.08380}{{\ttfamily
  2201.08380}}.

\bibitem{Montero:2021otb}
M.~Montero, C.~Vafa, T.~Van~Riet and G.~Venken, \emph{{The FL bound and its
  phenomenological implications}},
  \href{https://doi.org/10.1007/JHEP10(2021)009}{\emph{JHEP} {\bfseries 10}
  (2021) 009} [\href{https://arxiv.org/abs/2106.07650}{{\ttfamily
  2106.07650}}].

\bibitem{Dolan:2017vmn}
M.J.~Dolan, P.~Draper, J.~Kozaczuk and H.~Patel, \emph{{Transplanckian
  Censorship and Global Cosmic Strings}},
  \href{https://doi.org/10.1007/JHEP04(2017)133}{\emph{JHEP} {\bfseries 04}
  (2017) 133} [\href{https://arxiv.org/abs/1701.05572}{{\ttfamily
  1701.05572}}].

\bibitem{Hebecker:2017wsu}
A.~Hebecker, P.~Henkenjohann and L.T.~Witkowski, \emph{{What is the Magnetic
  Weak Gravity Conjecture for Axions?}},
  \href{https://doi.org/10.1002/prop.201700011}{\emph{Fortsch. Phys.}
  {\bfseries 65} (2017) 1700011}
  [\href{https://arxiv.org/abs/1701.06553}{{\ttfamily 1701.06553}}].

\bibitem{Heidenreich:2017sim}
B.~Heidenreich, M.~Reece and T.~Rudelius, \emph{{The Weak Gravity Conjecture
  and Emergence from an Ultraviolet Cutoff}},
  \href{https://doi.org/10.1140/epjc/s10052-018-5811-3}{\emph{Eur. Phys. J. C}
  {\bfseries 78} (2018) 337}
  [\href{https://arxiv.org/abs/1712.01868}{{\ttfamily 1712.01868}}].

\bibitem{Arkani-Hamed:2005zuc}
N.~Arkani-Hamed, S.~Dimopoulos and S.~Kachru, \emph{{Predictive landscapes and
  new physics at a TeV}},
  \href{https://arxiv.org/abs/hep-th/0501082}{{\ttfamily hep-th/0501082}}.

\bibitem{Distler:2005hi}
J.~Distler and U.~Varadarajan, \emph{{Random polynomials and the friendly
  landscape}},  \href{https://arxiv.org/abs/hep-th/0507090}{{\ttfamily
  hep-th/0507090}}.

\bibitem{Dimopoulos:2005ac}
S.~Dimopoulos, S.~Kachru, J.~McGreevy and J.G.~Wacker, \emph{{N-flation}},
  \href{https://doi.org/10.1088/1475-7516/2008/08/003}{\emph{JCAP} {\bfseries
  08} (2008) 003} [\href{https://arxiv.org/abs/hep-th/0507205}{{\ttfamily
  hep-th/0507205}}].

\bibitem{Dvali:2007hz}
G.~Dvali, \emph{{Black Holes and Large N Species Solution to the Hierarchy
  Problem}}, \href{https://doi.org/10.1002/prop.201000009}{\emph{Fortsch.
  Phys.} {\bfseries 58} (2010) 528}
  [\href{https://arxiv.org/abs/0706.2050}{{\ttfamily 0706.2050}}].

\bibitem{Holdom:1985ag}
B.~Holdom, \emph{{Two U(1)'s and Epsilon Charge Shifts}},
  \href{https://doi.org/10.1016/0370-2693(86)91377-8}{\emph{Phys. Lett. B}
  {\bfseries 166} (1986) 196}.

\bibitem{Contino:2020tix}
R.~Contino, K.~Max and R.K.~Mishra, \emph{{Searching for elusive dark sectors
  with terrestrial and celestial observations}},
  \href{https://doi.org/10.1007/JHEP06(2021)127}{\emph{JHEP} {\bfseries 06}
  (2021) 127} [\href{https://arxiv.org/abs/2012.08537}{{\ttfamily
  2012.08537}}].

\bibitem{Agrawal:2016quu}
P.~Agrawal, F.-Y.~Cyr-Racine, L.~Randall and J.~Scholtz, \emph{{Make Dark
  Matter Charged Again}},
  \href{https://doi.org/10.1088/1475-7516/2017/05/022}{\emph{JCAP} {\bfseries
  05} (2017) 022} [\href{https://arxiv.org/abs/1610.04611}{{\ttfamily
  1610.04611}}].

\bibitem{Berlin:2020pey}
A.~Berlin and A.~Hook, \emph{{Searching for Millicharged Particles with
  Superconducting Radio-Frequency Cavities}},
  \href{https://doi.org/10.1103/PhysRevD.102.035010}{\emph{Phys. Rev. D}
  {\bfseries 102} (2020) 035010}
  [\href{https://arxiv.org/abs/2001.02679}{{\ttfamily 2001.02679}}].

\bibitem{Polchinski:2003bq}
J.~Polchinski, \emph{{Monopoles, duality, and string theory}},
  \href{https://doi.org/10.1142/S0217751X0401866X}{\emph{Int. J. Mod. Phys. A}
  {\bfseries 19S1} (2004) 145}
  [\href{https://arxiv.org/abs/hep-th/0304042}{{\ttfamily hep-th/0304042}}].

\bibitem{Harlow:2018tng}
D.~Harlow and H.~Ooguri, \emph{{Symmetries in quantum field theory and quantum
  gravity}}, \href{https://doi.org/10.1007/s00220-021-04040-y}{\emph{Commun.
  Math. Phys.} {\bfseries 383} (2021) 1669}
  [\href{https://arxiv.org/abs/1810.05338}{{\ttfamily 1810.05338}}].

\bibitem{Cheung:2014vva}
C.~Cheung and G.N.~Remmen, \emph{{Naturalness and the Weak Gravity
  Conjecture}},
  \href{https://doi.org/10.1103/PhysRevLett.113.051601}{\emph{Phys. Rev. Lett.}
  {\bfseries 113} (2014) 051601}
  [\href{https://arxiv.org/abs/1402.2287}{{\ttfamily 1402.2287}}].

\bibitem{Cordova:2022rer}
C.~Cordova, K.~Ohmori and T.~Rudelius, \emph{{Generalized Symmetry Breaking
  Scales and Weak Gravity Conjectures}},
  \href{https://arxiv.org/abs/2202.05866}{{\ttfamily 2202.05866}}.

\bibitem{nariai1950some}
H.~Nariai, \emph{On some static solutions of einstein's gravitational field
  equations in a spherically symmetric case}, {\emph{Sci. Rep. Tohoku Univ.
  Eighth Ser.} {\bfseries 34} (1950) 160}.

\bibitem{Romans:1991nq}
L.J.~Romans, \emph{{Supersymmetric, cold and lukewarm black holes in
  cosmological Einstein-Maxwell theory}},
  \href{https://doi.org/10.1016/0550-3213(92)90684-4}{\emph{Nucl. Phys. B}
  {\bfseries 383} (1992) 395}
  [\href{https://arxiv.org/abs/hep-th/9203018}{{\ttfamily hep-th/9203018}}].

\bibitem{Hiscock:1990ex}
W.A.~Hiscock and L.D.~Weems, \emph{{Evolution of Charged Evaporating Black
  Holes}}, \href{https://doi.org/10.1103/PhysRevD.41.1142}{\emph{Phys. Rev. D}
  {\bfseries 41} (1990) 1142}.

\bibitem{Gibbons:1975kk}
G.W.~Gibbons, \emph{{Vacuum Polarization and the Spontaneous Loss of Charge by
  Black Holes}}, \href{https://doi.org/10.1007/BF01609829}{\emph{Commun. Math.
  Phys.} {\bfseries 44} (1975) 245}.

\bibitem{Schwinger:1951nm}
J.S.~Schwinger, \emph{{On gauge invariance and vacuum polarization}},
  \href{https://doi.org/10.1103/PhysRev.82.664}{\emph{Phys. Rev.} {\bfseries
  82} (1951) 664}.

\bibitem{Ahlers:2006iz}
M.~Ahlers, H.~Gies, J.~Jaeckel and A.~Ringwald, \emph{{On the Particle
  Interpretation of the PVLAS Data: Neutral versus Charged Particles}},
  \href{https://doi.org/10.1103/PhysRevD.75.035011}{\emph{Phys. Rev. D}
  {\bfseries 75} (2007) 035011}
  [\href{https://arxiv.org/abs/hep-ph/0612098}{{\ttfamily hep-ph/0612098}}].

\bibitem{DellaValle:2014xoa}
F.~Della~Valle, E.~Milotti, A.~Ejlli, G.~Messineo, L.~Piemontese, G.~Zavattini
  et~al., \emph{{First results from the new PVLAS apparatus: A new limit on
  vacuum magnetic birefringence}},
  \href{https://doi.org/10.1103/PhysRevD.90.092003}{\emph{Phys. Rev. D}
  {\bfseries 90} (2014) 092003}
  [\href{https://arxiv.org/abs/1406.6518}{{\ttfamily 1406.6518}}].

\bibitem{Vogel:2013raa}
H.~Vogel and J.~Redondo, \emph{{Dark Radiation constraints on minicharged
  particles in models with a hidden photon}},
  \href{https://doi.org/10.1088/1475-7516/2014/02/029}{\emph{JCAP} {\bfseries
  02} (2014) 029} [\href{https://arxiv.org/abs/1311.2600}{{\ttfamily
  1311.2600}}].

\bibitem{Jaeckel:2021xyo}
J.~Jaeckel and S.~Schenk, \emph{{Challenging the Stability of Light
  Millicharged Dark Matter}},
  \href{https://doi.org/10.1103/PhysRevD.103.103523}{\emph{Phys. Rev. D}
  {\bfseries 103} (2021) 103523}
  [\href{https://arxiv.org/abs/2102.08394}{{\ttfamily 2102.08394}}].

\bibitem{Ooguri:2006in}
H.~Ooguri and C.~Vafa, \emph{{On the Geometry of the String Landscape and the
  Swampland}},
  \href{https://doi.org/10.1016/j.nuclphysb.2006.10.033}{\emph{Nucl. Phys. B}
  {\bfseries 766} (2007) 21}
  [\href{https://arxiv.org/abs/hep-th/0605264}{{\ttfamily hep-th/0605264}}].

\bibitem{Grimm:2018ohb}
T.W.~Grimm, E.~Palti and I.~Valenzuela, \emph{{Infinite Distances in Field
  Space and Massless Towers of States}},
  \href{https://doi.org/10.1007/JHEP08(2018)143}{\emph{JHEP} {\bfseries 08}
  (2018) 143} [\href{https://arxiv.org/abs/1802.08264}{{\ttfamily
  1802.08264}}].

\bibitem{Prinz:1998ua}
A.A.~Prinz et~al., \emph{{Search for millicharged particles at SLAC}},
  \href{https://doi.org/10.1103/PhysRevLett.81.1175}{\emph{Phys. Rev. Lett.}
  {\bfseries 81} (1998) 1175}
  [\href{https://arxiv.org/abs/hep-ex/9804008}{{\ttfamily hep-ex/9804008}}].

\bibitem{SHIP:2021tpn}
{\scshape SHIP} collaboration, \emph{{The SHiP experiment at the proposed CERN
  SPS Beam Dump Facility}},  \href{https://arxiv.org/abs/2112.01487}{{\ttfamily
  2112.01487}}.

\bibitem{LDMX:2018cma}
{\scshape LDMX} collaboration, \emph{{Light Dark Matter eXperiment (LDMX)}},
  \href{https://arxiv.org/abs/1808.05219}{{\ttfamily 1808.05219}}.

\bibitem{Davidson:2000hf}
S.~Davidson, S.~Hannestad and G.~Raffelt, \emph{{Updated bounds on millicharged
  particles}}, \href{https://doi.org/10.1088/1126-6708/2000/05/003}{\emph{JHEP}
  {\bfseries 05} (2000) 003}
  [\href{https://arxiv.org/abs/hep-ph/0001179}{{\ttfamily hep-ph/0001179}}].

\bibitem{Lasenby:2020rlf}
R.~Lasenby, \emph{{Long range dark matter self-interactions and plasma
  instabilities}},
  \href{https://doi.org/10.1088/1475-7516/2020/11/034}{\emph{JCAP} {\bfseries
  11} (2020) 034} [\href{https://arxiv.org/abs/2007.00667}{{\ttfamily
  2007.00667}}].

\bibitem{Cruz:2022otv}
A.~Cruz and M.~McQuinn, \emph{{Astrophysical Plasma Instabilities induced by
  Long-Range Interacting Dark Matter}},
  \href{https://arxiv.org/abs/2202.12464}{{\ttfamily 2202.12464}}.

\bibitem{Kadota:2016tqq}
K.~Kadota, T.~Sekiguchi and H.~Tashiro, \emph{{A new constraint on millicharged
  dark matter from galaxy clusters}},
  \href{https://arxiv.org/abs/1602.04009}{{\ttfamily 1602.04009}}.

\bibitem{Stebbins:2019xjr}
A.~Stebbins and G.~Krnjaic, \emph{{New Limits on Charged Dark Matter from
  Large-Scale Coherent Magnetic Fields}},
  \href{https://doi.org/10.1088/1475-7516/2019/12/003}{\emph{JCAP} {\bfseries
  12} (2019) 003} [\href{https://arxiv.org/abs/1908.05275}{{\ttfamily
  1908.05275}}].

\bibitem{Munoz:2018pzp}
J.B.~Mu\~noz and A.~Loeb, \emph{{A small amount of mini-charged dark matter
  could cool the baryons in the early Universe}},
  \href{https://doi.org/10.1038/s41586-018-0151-x}{\emph{Nature} {\bfseries
  557} (2018) 684} [\href{https://arxiv.org/abs/1802.10094}{{\ttfamily
  1802.10094}}].

\bibitem{Caputo:2019tms}
A.~Caputo, L.~Sberna, M.~Frias, D.~Blas, P.~Pani, L.~Shao et~al.,
  \emph{{Constraints on millicharged dark matter and axionlike particles from
  timing of radio waves}},
  \href{https://doi.org/10.1103/PhysRevD.100.063515}{\emph{Phys. Rev. D}
  {\bfseries 100} (2019) 063515}
  [\href{https://arxiv.org/abs/1902.02695}{{\ttfamily 1902.02695}}].

\bibitem{Sikivie:2006ni}
P.~Sikivie, \emph{{Axion Cosmology}},
  \href{https://doi.org/10.1007/978-3-540-73518-2_2}{\emph{Lect. Notes Phys.}
  {\bfseries 741} (2008) 19}
  [\href{https://arxiv.org/abs/astro-ph/0610440}{{\ttfamily
  astro-ph/0610440}}].

\bibitem{Graham:2015rva}
P.W.~Graham, J.~Mardon and S.~Rajendran, \emph{{Vector Dark Matter from
  Inflationary Fluctuations}},
  \href{https://doi.org/10.1103/PhysRevD.93.103520}{\emph{Phys. Rev. D}
  {\bfseries 93} (2016) 103520}
  [\href{https://arxiv.org/abs/1504.02102}{{\ttfamily 1504.02102}}].

\bibitem{Co:2017mop}
R.T.~Co, L.J.~Hall and K.~Harigaya, \emph{{QCD Axion Dark Matter with a Small
  Decay Constant}},
  \href{https://doi.org/10.1103/PhysRevLett.120.211602}{\emph{Phys. Rev. Lett.}
  {\bfseries 120} (2018) 211602}
  [\href{https://arxiv.org/abs/1711.10486}{{\ttfamily 1711.10486}}].

\bibitem{Agrawal:2018vin}
P.~Agrawal, N.~Kitajima, M.~Reece, T.~Sekiguchi and F.~Takahashi, \emph{{Relic
  Abundance of Dark Photon Dark Matter}},
  \href{https://doi.org/10.1016/j.physletb.2019.135136}{\emph{Phys. Lett. B}
  {\bfseries 801} (2020) 135136}
  [\href{https://arxiv.org/abs/1810.07188}{{\ttfamily 1810.07188}}].

\bibitem{Fan:2013yva}
J.~Fan, A.~Katz, L.~Randall and M.~Reece, \emph{{Double-Disk Dark Matter}},
  \href{https://doi.org/10.1016/j.dark.2013.07.001}{\emph{Phys. Dark Univ.}
  {\bfseries 2} (2013) 139} [\href{https://arxiv.org/abs/1303.1521}{{\ttfamily
  1303.1521}}].

\bibitem{Fabbrichesi:2020wbt}
M.~Fabbrichesi, E.~Gabrielli and G.~Lanfranchi, \emph{{The Dark Photon}},
  \href{https://arxiv.org/abs/2005.01515}{{\ttfamily 2005.01515}}.

\bibitem{Reece:2009un}
M.~Reece and L.-T.~Wang, \emph{{Searching for the light dark gauge boson in
  GeV-scale experiments}},
  \href{https://doi.org/10.1088/1126-6708/2009/07/051}{\emph{JHEP} {\bfseries
  07} (2009) 051} [\href{https://arxiv.org/abs/0904.1743}{{\ttfamily
  0904.1743}}].

\bibitem{Co:2018lka}
R.T.~Co, A.~Pierce, Z.~Zhang and Y.~Zhao, \emph{{Dark Photon Dark Matter
  Produced by Axion Oscillations}},
  \href{https://doi.org/10.1103/PhysRevD.99.075002}{\emph{Phys. Rev. D}
  {\bfseries 99} (2019) 075002}
  [\href{https://arxiv.org/abs/1810.07196}{{\ttfamily 1810.07196}}].

\bibitem{Dror:2018pdh}
J.A.~Dror, K.~Harigaya and V.~Narayan, \emph{{Parametric Resonance Production
  of Ultralight Vector Dark Matter}},
  \href{https://doi.org/10.1103/PhysRevD.99.035036}{\emph{Phys. Rev. D}
  {\bfseries 99} (2019) 035036}
  [\href{https://arxiv.org/abs/1810.07195}{{\ttfamily 1810.07195}}].

\bibitem{Bastero-Gil:2018uel}
M.~Bastero-Gil, J.~Santiago, L.~Ubaldi and R.~Vega-Morales, \emph{{Vector dark
  matter production at the end of inflation}},
  \href{https://doi.org/10.1088/1475-7516/2019/04/015}{\emph{JCAP} {\bfseries
  04} (2019) 015} [\href{https://arxiv.org/abs/1810.07208}{{\ttfamily
  1810.07208}}].

\bibitem{Long:2019lwl}
A.J.~Long and L.-T.~Wang, \emph{{Dark Photon Dark Matter from a Network of
  Cosmic Strings}},
  \href{https://doi.org/10.1103/PhysRevD.99.063529}{\emph{Phys. Rev. D}
  {\bfseries 99} (2019) 063529}
  [\href{https://arxiv.org/abs/1901.03312}{{\ttfamily 1901.03312}}].

\bibitem{Bowman:2018yin}
J.D.~Bowman, A.E.E.~Rogers, R.A.~Monsalve, T.J.~Mozdzen and N.~Mahesh,
  \emph{{An absorption profile centred at 78 megahertz in the sky-averaged
  spectrum}}, \href{https://doi.org/10.1038/nature25792}{\emph{Nature}
  {\bfseries 555} (2018) 67}
  [\href{https://arxiv.org/abs/1810.05912}{{\ttfamily 1810.05912}}].

\bibitem{Pospelov:2018kdh}
M.~Pospelov, J.~Pradler, J.T.~Ruderman and A.~Urbano, \emph{{Room for New
  Physics in the Rayleigh-Jeans Tail of the Cosmic Microwave Background}},
  \href{https://doi.org/10.1103/PhysRevLett.121.031103}{\emph{Phys. Rev. Lett.}
  {\bfseries 121} (2018) 031103}
  [\href{https://arxiv.org/abs/1803.07048}{{\ttfamily 1803.07048}}].

\bibitem{Kovetz:2018zes}
E.D.~Kovetz, I.~Cholis and D.E.~Kaplan, \emph{{Bounds on ultralight
  hidden-photon dark matter from observation of the 21 cm signal at cosmic
  dawn}}, \href{https://doi.org/10.1103/PhysRevD.99.123511}{\emph{Phys. Rev. D}
  {\bfseries 99} (2019) 123511}
  [\href{https://arxiv.org/abs/1809.01139}{{\ttfamily 1809.01139}}].

\bibitem{Pospelov:2008zw}
M.~Pospelov, \emph{{Secluded U(1) below the weak scale}},
  \href{https://doi.org/10.1103/PhysRevD.80.095002}{\emph{Phys. Rev. D}
  {\bfseries 80} (2009) 095002}
  [\href{https://arxiv.org/abs/0811.1030}{{\ttfamily 0811.1030}}].

\bibitem{BaBar:2014zli}
{\scshape BaBar} collaboration, \emph{{Search for a Dark Photon in $e^+e^-$
  Collisions at BaBar}},
  \href{https://doi.org/10.1103/PhysRevLett.113.201801}{\emph{Phys. Rev. Lett.}
  {\bfseries 113} (2014) 201801}
  [\href{https://arxiv.org/abs/1406.2980}{{\ttfamily 1406.2980}}].

\bibitem{LHCb:2019vmc}
{\scshape LHCb} collaboration, \emph{{Search for $A'\to\mu^+\mu^-$ Decays}},
  \href{https://doi.org/10.1103/PhysRevLett.124.041801}{\emph{Phys. Rev. Lett.}
  {\bfseries 124} (2020) 041801}
  [\href{https://arxiv.org/abs/1910.06926}{{\ttfamily 1910.06926}}].

\bibitem{Bjorken:2009mm}
J.D.~Bjorken, R.~Essig, P.~Schuster and N.~Toro, \emph{{New Fixed-Target
  Experiments to Search for Dark Gauge Forces}},
  \href{https://doi.org/10.1103/PhysRevD.80.075018}{\emph{Phys. Rev. D}
  {\bfseries 80} (2009) 075018}
  [\href{https://arxiv.org/abs/0906.0580}{{\ttfamily 0906.0580}}].

\bibitem{Cesarotti:2022ttv}
C.~Cesarotti, S.~Homiller, R.K.~Mishra and M.~Reece, \emph{{Probing New Gauge
  Forces with a High-Energy Muon Beam Dump}},
  \href{https://arxiv.org/abs/2202.12302}{{\ttfamily 2202.12302}}.

\bibitem{Caputo:2021eaa}
A.~Caputo, A.J.~Millar, C.A.J.~O'Hare and E.~Vitagliano, \emph{{Dark photon
  limits: A handbook}},
  \href{https://doi.org/10.1103/PhysRevD.104.095029}{\emph{Phys. Rev. D}
  {\bfseries 104} (2021) 095029}
  [\href{https://arxiv.org/abs/2105.04565}{{\ttfamily 2105.04565}}].

\bibitem{Alvarez-Gaume:1986ghj}
L.~Alvarez-Gaume, P.H.~Ginsparg, G.W.~Moore and C.~Vafa, \emph{{An O(16) x
  O(16) Heterotic String}},
  \href{https://doi.org/10.1016/0370-2693(86)91524-8}{\emph{Phys. Lett. B}
  {\bfseries 171} (1986) 155}.

\bibitem{Pospelov:2008jk}
M.~Pospelov, A.~Ritz and M.B.~Voloshin, \emph{{Bosonic super-WIMPs as keV-scale
  dark matter}}, \href{https://doi.org/10.1103/PhysRevD.78.115012}{\emph{Phys.
  Rev. D} {\bfseries 78} (2008) 115012}
  [\href{https://arxiv.org/abs/0807.3279}{{\ttfamily 0807.3279}}].

\bibitem{Batell:2009yf}
B.~Batell, M.~Pospelov and A.~Ritz, \emph{{Probing a Secluded U(1) at
  B-factories}}, \href{https://doi.org/10.1103/PhysRevD.79.115008}{\emph{Phys.
  Rev. D} {\bfseries 79} (2009) 115008}
  [\href{https://arxiv.org/abs/0903.0363}{{\ttfamily 0903.0363}}].

\bibitem{An:2013yfc}
H.~An, M.~Pospelov and J.~Pradler, \emph{{New stellar constraints on dark
  photons}}, \href{https://doi.org/10.1016/j.physletb.2013.07.008}{\emph{Phys.
  Lett. B} {\bfseries 725} (2013) 190}
  [\href{https://arxiv.org/abs/1302.3884}{{\ttfamily 1302.3884}}].

\bibitem{Davidson:1991si}
S.~Davidson, B.~Campbell and D.C.~Bailey, \emph{{Limits on particles of small
  electric charge}},
  \href{https://doi.org/10.1103/PhysRevD.43.2314}{\emph{Phys. Rev. D}
  {\bfseries 43} (1991) 2314}.

\bibitem{Garcia:2020qrp}
A.A.~Garcia, K.~Bondarenko, S.~Ploeckinger, J.~Pradler and A.~Sokolenko,
  \emph{{Effective photon mass and (dark) photon conversion in the
  inhomogeneous Universe}},
  \href{https://doi.org/10.1088/1475-7516/2020/10/011}{\emph{JCAP} {\bfseries
  10} (2020) 011} [\href{https://arxiv.org/abs/2003.10465}{{\ttfamily
  2003.10465}}].

\bibitem{Caputo:2020bdy}
A.~Caputo, H.~Liu, S.~Mishra-Sharma and J.T.~Ruderman, \emph{{Dark Photon
  Oscillations in Our Inhomogeneous Universe}},
  \href{https://doi.org/10.1103/PhysRevLett.125.221303}{\emph{Phys. Rev. Lett.}
  {\bfseries 125} (2020) 221303}
  [\href{https://arxiv.org/abs/2002.05165}{{\ttfamily 2002.05165}}].

\bibitem{Chu:2011be}
X.~Chu, T.~Hambye and M.H.G.~Tytgat, \emph{{The Four Basic Ways of Creating
  Dark Matter Through a Portal}},
  \href{https://doi.org/10.1088/1475-7516/2012/05/034}{\emph{JCAP} {\bfseries
  05} (2012) 034} [\href{https://arxiv.org/abs/1112.0493}{{\ttfamily
  1112.0493}}].

\bibitem{Hall:2009bx}
L.J.~Hall, K.~Jedamzik, J.~March-Russell and S.M.~West, \emph{{Freeze-In
  Production of FIMP Dark Matter}},
  \href{https://doi.org/10.1007/JHEP03(2010)080}{\emph{JHEP} {\bfseries 03}
  (2010) 080} [\href{https://arxiv.org/abs/0911.1120}{{\ttfamily 0911.1120}}].

\bibitem{Bernal:2017kxu}
N.~Bernal, M.~Heikinheimo, T.~Tenkanen, K.~Tuominen and V.~Vaskonen, \emph{{The
  Dawn of FIMP Dark Matter: A Review of Models and Constraints}},
  \href{https://doi.org/10.1142/S0217751X1730023X}{\emph{Int. J. Mod. Phys. A}
  {\bfseries 32} (2017) 1730023}
  [\href{https://arxiv.org/abs/1706.07442}{{\ttfamily 1706.07442}}].

\bibitem{Dvorkin:2019zdi}
C.~Dvorkin, T.~Lin and K.~Schutz, \emph{{Making dark matter out of light:
  freeze-in from plasma effects}},
  \href{https://doi.org/10.1103/PhysRevD.99.115009}{\emph{Phys. Rev. D}
  {\bfseries 99} (2019) 115009}
  [\href{https://arxiv.org/abs/1902.08623}{{\ttfamily 1902.08623}}].

\bibitem{Fixsen:1996nj}
D.J.~Fixsen, E.S.~Cheng, J.M.~Gales, J.C.~Mather, R.A.~Shafer and E.L.~Wright,
  \emph{{The Cosmic Microwave Background spectrum from the full COBE FIRAS data
  set}}, \href{https://doi.org/10.1086/178173}{\emph{Astrophys. J.} {\bfseries
  473} (1996) 576} [\href{https://arxiv.org/abs/astro-ph/9605054}{{\ttfamily
  astro-ph/9605054}}].

\bibitem{Ehret:2010mh}
K.~Ehret et~al., \emph{{New ALPS Results on Hidden-Sector Lightweights}},
  \href{https://doi.org/10.1016/j.physletb.2010.04.066}{\emph{Phys. Lett. B}
  {\bfseries 689} (2010) 149}
  [\href{https://arxiv.org/abs/1004.1313}{{\ttfamily 1004.1313}}].

\bibitem{Inada:2013tx}
T.~Inada, T.~Namba, S.~Asai, T.~Kobayashi, Y.~Tanaka, K.~Tamasaku et~al.,
  \emph{{Results of a Search for Paraphotons with Intense X-ray Beams at
  SPring-8}}, \href{https://doi.org/10.1016/j.physletb.2013.04.033}{\emph{Phys.
  Lett. B} {\bfseries 722} (2013) 301}
  [\href{https://arxiv.org/abs/1301.6557}{{\ttfamily 1301.6557}}].

\bibitem{Povey:2010hs}
R.~Povey, J.~Hartnett and M.~Tobar, \emph{{Microwave cavity light shining
  through a wall optimization and experiment}},
  \href{https://doi.org/10.1103/PhysRevD.82.052003}{\emph{Phys. Rev. D}
  {\bfseries 82} (2010) 052003}
  [\href{https://arxiv.org/abs/1003.0964}{{\ttfamily 1003.0964}}].

\bibitem{Parker:2013fxa}
S.R.~Parker, J.G.~Hartnett, R.G.~Povey and M.E.~Tobar, \emph{{Cryogenic
  resonant microwave cavity searches for hidden sector photons}},
  \href{https://doi.org/10.1103/PhysRevD.88.112004}{\emph{Phys. Rev. D}
  {\bfseries 88} (2013) 112004}
  [\href{https://arxiv.org/abs/1410.5244}{{\ttfamily 1410.5244}}].

\bibitem{ADMX:2010ubl}
{\scshape ADMX} collaboration, \emph{{A Search for Hidden Sector Photons with
  ADMX}}, \href{https://doi.org/10.1103/PhysRevLett.105.171801}{\emph{Phys.
  Rev. Lett.} {\bfseries 105} (2010) 171801}
  [\href{https://arxiv.org/abs/1007.3766}{{\ttfamily 1007.3766}}].

\bibitem{Betz:2013dza}
M.~Betz, F.~Caspers, M.~Gasior, M.~Thumm and S.W.~Rieger, \emph{{First results
  of the CERN Resonant Weakly Interacting sub-eV Particle Search (CROWS)}},
  \href{https://doi.org/10.1103/PhysRevD.88.075014}{\emph{Phys. Rev. D}
  {\bfseries 88} (2013) 075014}
  [\href{https://arxiv.org/abs/1310.8098}{{\ttfamily 1310.8098}}].

\bibitem{Redondo:2013lna}
J.~Redondo and G.~Raffelt, \emph{{Solar constraints on hidden photons
  re-visited}},
  \href{https://doi.org/10.1088/1475-7516/2013/08/034}{\emph{JCAP} {\bfseries
  08} (2013) 034} [\href{https://arxiv.org/abs/1305.2920}{{\ttfamily
  1305.2920}}].

\bibitem{Lasenby:2020goo}
R.~Lasenby and K.~Van~Tilburg, \emph{{Dark photons in the solar basin}},
  \href{https://doi.org/10.1103/PhysRevD.104.023020}{\emph{Phys. Rev. D}
  {\bfseries 104} (2021) 023020}
  [\href{https://arxiv.org/abs/2008.08594}{{\ttfamily 2008.08594}}].

\bibitem{Cardoso:2018tly}
V.~Cardoso, O.J.C.~Dias, G.S.~Hartnett, M.~Middleton, P.~Pani and J.E.~Santos,
  \emph{{Constraining the mass of dark photons and axion-like particles through
  black-hole superradiance}},
  \href{https://doi.org/10.1088/1475-7516/2018/03/043}{\emph{JCAP} {\bfseries
  03} (2018) 043} [\href{https://arxiv.org/abs/1801.01420}{{\ttfamily
  1801.01420}}].

\bibitem{Buen-Abad:2015ova}
M.A.~Buen-Abad, G.~Marques-Tavares and M.~Schmaltz, \emph{{Non-Abelian dark
  matter and dark radiation}},
  \href{https://doi.org/10.1103/PhysRevD.92.023531}{\emph{Phys. Rev. D}
  {\bfseries 92} (2015) 023531}
  [\href{https://arxiv.org/abs/1505.03542}{{\ttfamily 1505.03542}}].

\bibitem{Verde:2019ivm}
L.~Verde, T.~Treu and A.G.~Riess, \emph{{Tensions between the Early and the
  Late Universe}},
  \href{https://doi.org/10.1038/s41550-019-0902-0}{\emph{Nature Astron.}
  {\bfseries 3} (2019) 891} [\href{https://arxiv.org/abs/1907.10625}{{\ttfamily
  1907.10625}}].

\bibitem{Chen:2009ab}
F.~Chen, J.M.~Cline and A.R.~Frey, \emph{{Nonabelian dark matter: Models and
  constraints}}, \href{https://doi.org/10.1103/PhysRevD.80.083516}{\emph{Phys.
  Rev. D} {\bfseries 80} (2009) 083516}
  [\href{https://arxiv.org/abs/0907.4746}{{\ttfamily 0907.4746}}].

\bibitem{Battye:2014qga}
R.A.~Battye, T.~Charnock and A.~Moss, \emph{{Tension between the power spectrum
  of density perturbations measured on large and small scales}},
  \href{https://doi.org/10.1103/PhysRevD.91.103508}{\emph{Phys. Rev. D}
  {\bfseries 91} (2015) 103508}
  [\href{https://arxiv.org/abs/1409.2769}{{\ttfamily 1409.2769}}].

\bibitem{Banks:2010zn}
T.~Banks and N.~Seiberg, \emph{{Symmetries and Strings in Field Theory and
  Gravity}}, \href{https://doi.org/10.1103/PhysRevD.83.084019}{\emph{Phys. Rev.
  D} {\bfseries 83} (2011) 084019}
  [\href{https://arxiv.org/abs/1011.5120}{{\ttfamily 1011.5120}}].

\bibitem{Ahlers:2008qc}
M.~Ahlers, J.~Jaeckel, J.~Redondo and A.~Ringwald, \emph{{Probing Hidden Sector
  Photons through the Higgs Window}},
  \href{https://doi.org/10.1103/PhysRevD.78.075005}{\emph{Phys. Rev. D}
  {\bfseries 78} (2008) 075005}
  [\href{https://arxiv.org/abs/0807.4143}{{\ttfamily 0807.4143}}].

\bibitem{An:2013yua}
H.~An, M.~Pospelov and J.~Pradler, \emph{{Dark Matter Detectors as Dark Photon
  Helioscopes}},
  \href{https://doi.org/10.1103/PhysRevLett.111.041302}{\emph{Phys. Rev. Lett.}
  {\bfseries 111} (2013) 041302}
  [\href{https://arxiv.org/abs/1304.3461}{{\ttfamily 1304.3461}}].

\bibitem{Lee:2021cor}
S.M.~Lee, D.Y.~Cheong, S.C.~Hyun, S.C.~Park and M.-S.~Seo, \emph{{Festina-Lente
  bound on Higgs vacuum structure and inflation}},
  \href{https://doi.org/10.1007/JHEP02(2022)100}{\emph{JHEP} {\bfseries 02}
  (2022) 100} [\href{https://arxiv.org/abs/2111.04010}{{\ttfamily
  2111.04010}}].

\bibitem{Guidetti:2022xct}
V.~Guidetti, N.~Righi, G.~Venken and A.~Westphal, \emph{{Axionic Festina
  Lente}},  \href{https://arxiv.org/abs/2206.03494}{{\ttfamily 2206.03494}}.

\bibitem{Shiu:2013wxa}
G.~Shiu, P.~Soler and F.~Ye, \emph{{Milli-Charged Dark Matter in Quantum
  Gravity and String Theory}},
  \href{https://doi.org/10.1103/PhysRevLett.110.241304}{\emph{Phys. Rev. Lett.}
  {\bfseries 110} (2013) 241304}
  [\href{https://arxiv.org/abs/1302.5471}{{\ttfamily 1302.5471}}].

\bibitem{McDonough:2021pdg}
E.~McDonough, M.-X.~Lin, J.C.~Hill, W.~Hu and S.~Zhou, \emph{{The Early Dark
  Sector, the Hubble Tension, and the Swampland}},
  \href{https://arxiv.org/abs/2112.09128}{{\ttfamily 2112.09128}}.

\bibitem{Heidenreich:2016aqi}
B.~Heidenreich, M.~Reece and T.~Rudelius, \emph{{Evidence for a sublattice weak
  gravity conjecture}},
  \href{https://doi.org/10.1007/JHEP08(2017)025}{\emph{JHEP} {\bfseries 08}
  (2017) 025} [\href{https://arxiv.org/abs/1606.08437}{{\ttfamily
  1606.08437}}].

\bibitem{Noumi:2022zht}
T.~Noumi, S.~Sato and J.~Tokuda, \emph{{Phenomenological Motivation for
  Gravitational Positivity Bounds: A Case Study of Dark Sector Physics}},
  \href{https://arxiv.org/abs/2205.12835}{{\ttfamily 2205.12835}}.

\bibitem{Maleknejad:2011jw}
A.~Maleknejad and M.M.~Sheikh-Jabbari, \emph{{Gauge-flation: Inflation From
  Non-Abelian Gauge Fields}},
  \href{https://doi.org/10.1016/j.physletb.2013.05.001}{\emph{Phys. Lett. B}
  {\bfseries 723} (2013) 224}
  [\href{https://arxiv.org/abs/1102.1513}{{\ttfamily 1102.1513}}].

\bibitem{Kofman:2004yc}
L.~Kofman, A.D.~Linde, X.~Liu, A.~Maloney, L.~McAllister and E.~Silverstein,
  \emph{{Beauty is attractive: Moduli trapping at enhanced symmetry points}},
  \href{https://doi.org/10.1088/1126-6708/2004/05/030}{\emph{JHEP} {\bfseries
  05} (2004) 030} [\href{https://arxiv.org/abs/hep-th/0403001}{{\ttfamily
  hep-th/0403001}}].

\bibitem{Novikov:1985rd}
V.A.~Novikov, M.A.~Shifman, A.I.~Vainshtein and V.I.~Zakharov, \emph{{The beta
  function in supersymmetric gauge theories. Instantons versus traditional
  approach}}, \href{https://doi.org/10.1016/0370-2693(86)90810-5}{\emph{Phys.
  Lett. B} {\bfseries 166} (1986) 329}.

\bibitem{MishraNEW}
R.~Mishra, \emph{{To appear}}, .

\bibitem{Adshead:2012kp}
P.~Adshead and M.~Wyman, \emph{{Chromo-Natural Inflation: Natural inflation on
  a steep potential with classical non-Abelian gauge fields}},
  \href{https://doi.org/10.1103/PhysRevLett.108.261302}{\emph{Phys. Rev. Lett.}
  {\bfseries 108} (2012) 261302}
  [\href{https://arxiv.org/abs/1202.2366}{{\ttfamily 1202.2366}}].

\bibitem{Freese:1990rb}
K.~Freese, J.A.~Frieman and A.V.~Olinto, \emph{{Natural inflation with pseudo -
  Nambu-Goldstone bosons}},
  \href{https://doi.org/10.1103/PhysRevLett.65.3233}{\emph{Phys. Rev. Lett.}
  {\bfseries 65} (1990) 3233}.

\bibitem{Adams:1992bn}
F.C.~Adams, J.R.~Bond, K.~Freese, J.A.~Frieman and A.V.~Olinto, \emph{{Natural
  inflation: Particle physics models, power law spectra for large scale
  structure, and constraints from COBE}},
  \href{https://doi.org/10.1103/PhysRevD.47.426}{\emph{Phys. Rev. D} {\bfseries
  47} (1993) 426} [\href{https://arxiv.org/abs/hep-ph/9207245}{{\ttfamily
  hep-ph/9207245}}].

\bibitem{Planck:2018vyg}
{\scshape Planck} collaboration, \emph{{Planck 2018 results. VI. Cosmological
  parameters}},
  \href{https://doi.org/10.1051/0004-6361/201833910}{\emph{Astron. Astrophys.}
  {\bfseries 641} (2020) A6}
  [\href{https://arxiv.org/abs/1807.06209}{{\ttfamily 1807.06209}}].

\bibitem{BICEP2:2018kqh}
{\scshape BICEP2, Keck Array} collaboration, \emph{{BICEP2 / Keck Array x:
  Constraints on Primordial Gravitational Waves using Planck, WMAP, and New
  BICEP2/Keck Observations through the 2015 Season}},
  \href{https://doi.org/10.1103/PhysRevLett.121.221301}{\emph{Phys. Rev. Lett.}
  {\bfseries 121} (2018) 221301}
  [\href{https://arxiv.org/abs/1810.05216}{{\ttfamily 1810.05216}}].

\bibitem{delaFuente:2014aca}
A.~de~la Fuente, P.~Saraswat and R.~Sundrum, \emph{{Natural Inflation and
  Quantum Gravity}},
  \href{https://doi.org/10.1103/PhysRevLett.114.151303}{\emph{Phys. Rev. Lett.}
  {\bfseries 114} (2015) 151303}
  [\href{https://arxiv.org/abs/1412.3457}{{\ttfamily 1412.3457}}].

\bibitem{Rudelius:2014wla}
T.~Rudelius, \emph{{On the Possibility of Large Axion Moduli Spaces}},
  \href{https://doi.org/10.1088/1475-7516/2015/04/049}{\emph{JCAP} {\bfseries
  1504} (2015) 049} [\href{https://arxiv.org/abs/1409.5793}{{\ttfamily
  1409.5793}}].

\bibitem{Rudelius:2015xta}
T.~Rudelius, \emph{{Constraints on Axion Inflation from the Weak Gravity
  Conjecture}}, \href{https://doi.org/10.1088/1475-7516/2015/09/020,
  10.1088/1475-7516/2015/9/020}{\emph{JCAP} {\bfseries 1509} (2015) 020}
  [\href{https://arxiv.org/abs/1503.00795}{{\ttfamily 1503.00795}}].

\bibitem{Montero:2015ofa}
M.~Montero, A.M.~Uranga and I.~Valenzuela, \emph{{Transplanckian axions!?}},
  \href{https://doi.org/10.1007/JHEP08(2015)032}{\emph{JHEP} {\bfseries 08}
  (2015) 032} [\href{https://arxiv.org/abs/1503.03886}{{\ttfamily
  1503.03886}}].

\bibitem{Brown:2015iha}
J.~Brown, W.~Cottrell, G.~Shiu and P.~Soler, \emph{{Fencing in the Swampland:
  Quantum Gravity Constraints on Large Field Inflation}},
  \href{https://doi.org/10.1007/JHEP10(2015)023}{\emph{JHEP} {\bfseries 10}
  (2015) 023} [\href{https://arxiv.org/abs/1503.04783}{{\ttfamily
  1503.04783}}].

\bibitem{Bachlechner:2015qja}
T.C.~Bachlechner, C.~Long and L.~McAllister, \emph{{Planckian Axions and the
  Weak Gravity Conjecture}},
  \href{https://arxiv.org/abs/1503.07853}{{\ttfamily 1503.07853}}.

\bibitem{Hebecker:2015rya}
A.~Hebecker, P.~Mangat, F.~Rompineve and L.T.~Witkowski, \emph{{Winding out of
  the Swamp: Evading the Weak Gravity Conjecture with F-term Winding
  Inflation?}},
  \href{https://doi.org/10.1016/j.physletb.2015.07.026}{\emph{Phys. Lett.}
  {\bfseries B748} (2015) 455}
  [\href{https://arxiv.org/abs/1503.07912}{{\ttfamily 1503.07912}}].

\bibitem{Brown:2015lia}
J.~Brown, W.~Cottrell, G.~Shiu and P.~Soler, \emph{{On Axionic Field Ranges,
  Loopholes and the Weak Gravity Conjecture}},
  \href{https://arxiv.org/abs/1504.00659}{{\ttfamily 1504.00659}}.

\bibitem{Junghans:2015hba}
D.~Junghans, \emph{{Large-Field Inflation with Multiple Axions and the Weak
  Gravity Conjecture}},  \href{https://arxiv.org/abs/1504.03566}{{\ttfamily
  1504.03566}}.

\bibitem{Heidenreich:2015wga}
B.~Heidenreich, M.~Reece and T.~Rudelius, \emph{{Weak Gravity Strongly
  Constrains Large-Field Axion Inflation}},
  \href{https://arxiv.org/abs/1506.03447}{{\ttfamily 1506.03447}}.

\bibitem{Palti:2015xra}
E.~Palti, \emph{{On Natural Inflation and Moduli Stabilisation in String
  Theory}}, \href{https://doi.org/10.1007/JHEP10(2015)188}{\emph{JHEP}
  {\bfseries 10} (2015) 188}
  [\href{https://arxiv.org/abs/1508.00009}{{\ttfamily 1508.00009}}].

\bibitem{Heidenreich:2015nta}
B.~Heidenreich, M.~Reece and T.~Rudelius, \emph{{Sharpening the Weak Gravity
  Conjecture with Dimensional Reduction}},
  \href{https://arxiv.org/abs/1509.06374}{{\ttfamily 1509.06374}}.

\bibitem{Kooner:2015rza}
K.~Kooner, S.~Parameswaran and I.~Zavala, \emph{{Warping the Weak Gravity
  Conjecture}},  \href{https://arxiv.org/abs/1509.07049}{{\ttfamily
  1509.07049}}.

\bibitem{Kappl:2015esy}
R.~Kappl, H.P.~Nilles and M.W.~Winkler, \emph{{Modulated Natural Inflation}},
  \href{https://arxiv.org/abs/1511.05560}{{\ttfamily 1511.05560}}.

\bibitem{Choi:2015aem}
K.~Choi and H.~Kim, \emph{{Aligned natural inflation with modulations}},
  \href{https://arxiv.org/abs/1511.07201}{{\ttfamily 1511.07201}}.

\bibitem{Adshead:2013qp}
P.~Adshead, E.~Martinec and M.~Wyman, \emph{{Gauge fields and inflation: Chiral
  gravitational waves, fluctuations, and the Lyth bound}},
  \href{https://doi.org/10.1103/PhysRevD.88.021302}{\emph{Phys. Rev. D}
  {\bfseries 88} (2013) 021302}
  [\href{https://arxiv.org/abs/1301.2598}{{\ttfamily 1301.2598}}].

\bibitem{Anber:2012du}
M.M.~Anber and L.~Sorbo, \emph{{Non-Gaussianities and chiral gravitational
  waves in natural steep inflation}},
  \href{https://doi.org/10.1103/PhysRevD.85.123537}{\emph{Phys. Rev. D}
  {\bfseries 85} (2012) 123537}
  [\href{https://arxiv.org/abs/1203.5849}{{\ttfamily 1203.5849}}].

\bibitem{Ji:2021mvg}
L.~Ji, D.E.~Kaplan, S.~Rajendran and E.H.~Tanin, \emph{{Thermal perturbations
  from cosmological constant relaxation}},
  \href{https://doi.org/10.1103/PhysRevD.105.015025}{\emph{Phys. Rev. D}
  {\bfseries 105} (2022) 015025}
  [\href{https://arxiv.org/abs/2109.05285}{{\ttfamily 2109.05285}}].

\bibitem{Graham:2015cka}
P.W.~Graham, D.E.~Kaplan and S.~Rajendran, \emph{{Cosmological Relaxation of
  the Electroweak Scale}},
  \href{https://doi.org/10.1103/PhysRevLett.115.221801}{\emph{Phys. Rev. Lett.}
  {\bfseries 115} (2015) 221801}
  [\href{https://arxiv.org/abs/1504.07551}{{\ttfamily 1504.07551}}].

\bibitem{Ibanez:2015fcv}
L.E.~Ibanez, M.~Montero, A.~Uranga and I.~Valenzuela, \emph{{Relaxion Monodromy
  and the Weak Gravity Conjecture}},
  \href{https://doi.org/10.1007/JHEP04(2016)020}{\emph{JHEP} {\bfseries 04}
  (2016) 020} [\href{https://arxiv.org/abs/1512.00025}{{\ttfamily
  1512.00025}}].

\bibitem{Hebecker:2015zss}
A.~Hebecker, F.~Rompineve and A.~Westphal, \emph{{Axion Monodromy and the Weak
  Gravity Conjecture}},
  \href{https://doi.org/10.1007/JHEP04(2016)157}{\emph{JHEP} {\bfseries 04}
  (2016) 157} [\href{https://arxiv.org/abs/1512.03768}{{\ttfamily
  1512.03768}}].

\end{thebibliography}\endgroup
\bibliographystyle{JHEP}

\begin{appendices}

\section{Festina Lente and Inflationary Models}
\label{app:FLandInflation}
In this appendix, we briefly comment on the constraints that the FL bound poses on existing inflationary scenarios (see \cite{Maleknejad:2011jw,Guidetti:2022xct} for related work).  The first is whether FL really applies during inflation, since the dS phase only lasts a finite amount of time, roughly 60 efolds.  In principle, one  could avoid FL if the charged Nariai black holes did not have enough time to discharge before inflation ends. As explained in Ref.~\cite{Montero:2019ekk}, the black-hole decay time is set by $t_{\rm bh}^{-1} \sim \sqrt{E} \sim \sqrt{g M_{\rm Pl} H}$. Demanding that this is longer than $N_e$ efolds means
  \begin{align*}
    \frac{1}{\sqrt{g M_{\rm Pl} H}} > \frac{N_e}{H} \implies g \lesssim \frac{H}{N_e^2 M_\mathrm{Pl}}.
  \end{align*}
  This limit is incompatible with the magnetic version of FL discussed in \cite{Montero:2021otb}. Hence, discussion of FL in dS is unavoidable.

Another potential loophole to avoid the FL discussion would be to exploit the $\mathcal{O}(H)$ thermal masses that particles get by virtue of their presence in a dS background. With masses of this order we have:
\begin{align}
    m^2 \sim H^2 \lesssim g M_\mathrm{Pl} H \implies H \lesssim g M_\mathrm{Pl}
    \label{eq:magneticFL}
\end{align}
again running afoul of the magnetic version of FL from~\cite{Montero:2021otb}. Another way to interpret the magnetic FL inequality~\eqref{eq:magneticFL} is that the gauge coupling in dS space cannot be made small enough to allow for thermal masses to satisfy the FL bound.

Since FL is unavoidable, we must study the fate of the Standard Model during inflation. Assuming a high-enough energy scale for inflation with the SM unchanged is problematic since the charged particles in the SM would violate the FL bound. This applies for particles like the electron but also for the non-Abelian gauge bosons.

Leptons and quarks can be made massive e.g.~by appealing to a model of Higgs inflation where the Higgs VEV is very large during the inflationary era, as discussed in \cite{Montero:2019ekk}. In addition, this has the advantage of Higgsing the $SU(2)$ gauge symmetry so that its gluons can be made sufficiently massive as well. In case the inflaton is not the Higgs, then some model-building effort is required. For instance, one could attempt to make the gauge couplings small by coupling the inflaton to the kinetic terms of the gauge fields so as to satisfy the FL bound. Here, however, we must be careful with quantum gravity cutoff scales. As we have seen above, the magnetic WGC implies that $2g^2 \geq 3 (H/M_{\rm Pl})^2$ meaning that the right hand side of the FL inequality must be greater than $3H^2$. Since particles are expected to get a mass of order Hubble during inflation, FL can potentially be marginally satisfied.
The QCD sector could also be Higgsed during inflation, albeit with a different field to the SM one. This might seem tuned at first but could be realized along the lines discussed in~\cite{Kofman:2004yc}.

Another possibility is that the non-Abelian sector (both $SU(3)$ and electroweak $SU(2)$ fields) is confined during inflation. This can happen naturally in the context of some Grand Unified Theories (GUT)'s. If their coupling takes the value $\alpha_0$ at an energy scale $\Lambda_0$, the confinement scale of a supersymmetric GUT can be determined via the NSVZ formula \cite{Novikov:1985rd},
\begin{equation}\Lambda= \Lambda_0 e^{-\frac{2\pi}{\alpha K}},\quad K\equiv 3 T_{Adj} - \sum_i T(R_i)\end{equation}
where the coefficients $T_{Adj}$ are group-theoretic. For $E_6$ grand unification, setting $\alpha_0$ to its grand-unification value at the scale $\Lambda_0\sim 10^{15}$ GeV gives $\Lambda\sim 10^{13}$ GeV, so that gauge fields would be confined during inflation (for $H_{\rm inf}$ above that scale). In this case, FL would apply automatically, but one would still be pressed to explain reheating in this model (which would be a strongly coupled process), and the setup could also have a monopole problem. We just present it as an illustration of the fact that, while the SM+inflation scenario is certainly incompatible with FL, the ways out are manifold.

More generally, any inflationary scenario that requires the use of non-Abelian gauge fields is incompatible with FL if the bound can be applied. As we have seen above, FL cannot be avoided by tuning the gauge coupling. That said, one sure way of avoiding FL is to have the lifetime of dS be shorter than the decay time of the Nariai black hole\footnote{We are grateful to Rashmish Mishra for insight and discussion about this point. For more details see~\cite{MishraNEW}.} If this condition is satisfied then FL cannot be applied. In fact, if one starts with an initial condition that is the dS Nariai black hole, then the cosmological constant would change considerably before the black hole has a chance to decay into the superextremal region. The conclusion of~\cite{Montero:2019ekk} cannot then be applied directly to this case and a more careful analysis is required.

Inflationary models where the FL bound may apply include chromo-natural inflation~\cite{Adshead:2012kp} and gauge-flation \cite{Maleknejad:2011jw} for example. In this appendix, we focus on the former. Chromo-natural inflation was proposed as an extension of natural inflation \cite{Freese:1990rb,Adams:1992bn} in order to circumvent the need for a super-Planckian axion decay constant. In order to match CMB observations~\cite{Planck:2018vyg,BICEP2:2018kqh}, the axion decay constant has to be super-Planckian, $f \gtrsim 5 M_{\rm Pl}$ something that has been argued extensively to be in the Swampland \cite{delaFuente:2014aca,Rudelius:2014wla,Rudelius:2015xta,Montero:2015ofa,Brown:2015iha,Bachlechner:2015qja,Hebecker:2015rya,Brown:2015lia,Junghans:2015hba,Heidenreich:2015wga,Palti:2015xra,Heidenreich:2015nta,Kooner:2015rza,Kappl:2015esy,Choi:2015aem}. Chromo-natural inflation avoids this problem by coupling the axion to a non-Abelian gauge field allowing for inflation on a steeper potential, i.e. with $f < M_{\rm Pl}$, allowing the axion's kinetic energy to be dissipated into a bath of non-Abelian gauge fields.

In this model, with $f < M_\mathrm{Pl}$, accelerated expansion without the non-Abelian background lasts for a relatively short time. To check the applicability of FL, we must then compare this time-scale to the black hole lifetime. For an axion near the top of a cosine potential of the form:
\begin{align*}
    V(\phi) \sim \Lambda^4 \cos \left(\frac{\phi}{f}\right),
\end{align*}
the condition for FL to apply, $t_\mathrm{dS} \gtrsim t_\mathrm{bh}$, translates to:
\begin{align}
    \label{eq:FLandCNI}
    \sqrt{g} f^2 \gtrsim \Lambda M_\mathrm{Pl}.
\end{align}
For the parameter values given the original paper~\cite{Adshead:2012kp}, this condition is not satisfied and thus it seems like the na\"{i}ve application of FL does not go through. That said, the inequality derived above~\ref{eq:FLandCNI}

Finally, the production of chiral gravitational-wave signals relying on non-Abelian gauge fields, such as in~\cite{Adshead:2013qp} have also to be considered carefully in light of the FL bound. This should motivate the search for other means of generating parity-violating gravitational-wave signals, such as those studied in~\cite{Anber:2012du}.

\section{Cosmological Constant Relaxation}
\label{app:CCrelaxion}

The model of~\cite{Ji:2021mvg}  attempts to provide a solution to the cosmological constant (CC) problem using a scalar field that dynamically relaxes the CC to a small value. This is similar in spirit to the original relaxion proposal for the EW hierarchy problem in \cite{Graham:2015cka} (see \cite{Ibanez:2015fcv,Hebecker:2015zss} for Swampland constraints on this scenario).

In this model, the relaxion field $\phi$ has a linear potential and is axially coupled to a dark $SU(N)_d$
gauge group, via the Lagrangian
\begin{align*}
  \mathcal{L} = \frac12 (\partial \phi)^2 +g_\phi^3\phi
  -\frac14 G_{\mu\nu}^a G^{a\mu\nu}
  -\frac{\alpha}{8\pi}\frac{\phi}{f_G} G_{\mu\nu}^a \tilde{G}^{a\mu\nu}.
\end{align*}
During the initial stages of evolution, the universe is in a phase of cold inflation. Soon after, a tachyonic instability leads to the sub-horizon production and amplification of the non-Abelian gauge fields which thermalize when their energy density exceeds a critical value. This leads to a steady state solution that persists for a long time and during which the gauge fields are continually sourced by the rolling relaxion and diluted by the Hubble expansion. Inflation ends shortly before the relaxion field crosses into a region where the potential is negative ($\phi > 0$ in the above Lagrangian). The relaxion continues rolling and comes to a halt at a negative value of the potential shortly afterwards. This results in a crunching universe which will have a growing energy density and excite UV degrees of freedom that lead to a bounce changing the sign of the CC due to contributions from other sectors not described by the above Lagrangian. After the bounce the universe enters the usual Hot Big Bang phase. More details can be found in the original paper~\cite{Ji:2021mvg}.

Thermal fluctuations in the  non-Abelian sector are essential for the viability of this scenario, since they produce inhomogeneities with the correct amplitude to match observations. Inhomogeneities produced by $\phi$ fluctuations are too small to account for observations. However, it is precisely these non-Abelian fields that cause the model to run afoul of the FL bound during the inflationary phase
as they ought to be unconfined. It is conceivable that another form of radiation that can thermalize during inflation can act as a substitute for the dark gluon bath but care must be taken to ensure that the masses of the constituent particles and the gauge interactions between them do not contradict the FL bound.

It is likely that the features of this model which make it incompatible with FL are not essential. A possible way out of our constraints would be to replace the non-Abelian gauge fields in the model by Abelian ones or, if strong interactions are necessary, by coupling to a Conformal Field Theory. The Swampland viewpoint often encourages exploration of phenomenological scenarios that do not fit in the standard notion of naturalness for a low-energy observer.

\end{appendices}

\end{document}